\documentclass[twocolumn,showpacs,preprintnumbers,amssymb,nofootinbib]{revtex4}

\usepackage{graphicx}
\usepackage{bm} 

\begin{document}


\title{Pair creation of de Sitter black holes on a cosmic string background}

\author{\'Oscar J. C. Dias}
\email{oscar@fisica.ist.utl.pt}
\author{Jos\'e P. S. Lemos}
\email{lemos@kelvin.ist.utl.pt} \affiliation{ Centro
Multidisciplinar de Astrof\'{\i}sica - CENTRA, Departamento de
F\'{\i}sica, Instituto Superior T\'ecnico, Av. Rovisco Pais 1,
1049-001 Lisbon}

\date{\today}
\begin{abstract}
We analyze the quantum process in which a cosmic string breaks in
a de Sitter (dS) background, and a pair of neutral or charged
black holes is produced at the ends of the string. The energy to
materialize and accelerate the pair comes from the positive
cosmological constant and, in addition, from the string tension.
The compact saddle point solutions without conical singularities
(instantons) or with  conical singularities (sub-maximal
instantons) that describe this process are constructed through the
analytical continuation of the dS C-metric. Then, we explicitly
compute the pair creation rate of the process. In particular, we
find the nucleation rate of a cosmic string in a dS background,
and the probability that it breaks and a pair of black holes is
produced. Finally we verify that, as occurs with pair production
processes in other background fields, the pair creation rate of
black holes is proportional to $e^{S}$, where the gravitational
entropy of the black hole, $S$, is given by one quarter of the
area of the horizons present in the saddle point solution that
mediates the process.
\end{abstract}

\pacs{04.70.Dy, 04.70.-s, 04.20.Gz, 98.80.Jk, 98.80.Cq}


\maketitle

\section{\label{sec:Int}Introduction}

In nature there are few known processes that allow the production
of black holes. The best well-known is the gravitational collapse
of a massive star or cluster of stars. Due to fermionic degeneracy
pressure these black holes cannot have a mass below the
Oppenheimer-Volkoff  limiting mass ($\sim 3 M_{\odot}$ in recent
calculations). Another one is the quantum Schwinger-like process
of black hole pair creation in an external field. These black
holes can have Planck dimensions and thus their evolution is ruled
by quantum effects. Moreover, gravitational pair creation involves
topology changing processes, and allows a study of the statistical
properties of black holes, namely: it favors the conjecture that
the number of internal microstates of a black hole is given by the
exponential of one-quarter of the area of the black hole horizon,
and it gives useful clues to the black hole information paradox.

The evaluation of the black hole pair creation rate has been done
at the semiclassical level using the instanton method.  An
instanton is an Euclidean solution that interpolates between the
initial and final states of a classically forbidden transition,
and is a saddle point for the Euclidean path integral that
describes the pair creation rate. This instanton method has been
first introduced in studies about decay of metastable
termodynamical states, and it has been applied in the context of
pair creation of particles and fields in the absence of gravity,
by several authors (see, e.g., \cite{OscLem_FalVac} for a review).

\subsection{\label{sec:Int-1}Backgrounds for black hole pair creation. Instantons}

The instanton method has been also introduced as a framework for
quantum gravity, with successful results in the analysis of
gravitational thermodynamic issues and black hole pair production
processes, among others (see \cite{EQG-book}). The regular
instantons that describe the process we are interested in - the
pair creation of black holes in an external field - can be
obtained by analytically continuing (i) a solution found by Ernst
\cite{Ernst}, (ii) the de Sitter black hole solutions, (iii) a
solution found by Kinnersley and Walker known as the C-metric
\cite{KW}, (iv) a combination of the above solutions, or (v) the
domain wall solution \cite{VilenkinStringIpserSikivie}. To each
one of these five families of instantons corresponds a different
way by which energy can be furnished in order to materialize the
pair of black holes and to accelerate them apart. In case (i) the
energy is provided by the electromagnetic Lorentz force, in case
(ii) the strings tension furnishes this energy, in case (iii) the
energy is provided by the rapid cosmological expansion associated
to the positive cosmological constant $\Lambda$, in case (iv) the
energy is provided by a combination of the above fields, and
finally in case (v) the energy is given by the repulsive
gravitational field of the domain wall. Since these solutions play
a fundamental role, we will now briefly discuss some of them that
are less known. The C-metric \cite{KW} describes a pair of black
holes (neutral or charged) uniformly accelerating in opposite
directions. The solution has conical singularities at its angular
poles that, when conveniently treated, can be interpreted as two
strings from each one of the black holes towards the infinity and
whose tension provides the necessary force to pull apart the black
holes. By appending a suitable external electromagnetic field,
Ernst \cite{Ernst} has removed all the conical singularities of
the charged flat C-metric. The Ernst solution then describes two
oppositely charged black holes undergoing uniform acceleration
provided by the Lorentz force associated to the external field
(for the magnetic solution see \cite{Ernst}, while the explicit
electric solution can be found in Brown \cite{Brown}).
Asymptotically, the Ernst solution reduces to the
 Melvin universe \cite{Melvin}. The Lorentz
sector of the C-metric and Ernst solution describe the evolution
of the black holes after their creation. The usual de Sitter black
hole solutions, when euclideanized, give also instantons for pair
creation of black holes. Indeed, the de Sitter black holes
solutions can be interpreted as representing a black hole pair
being accelerated by the cosmological constant.

It was believed that the only black hole pairs that could be
nucleated were those whose Euclidean sector was free of conical
singularities (instantons). This regularity condition restricted
the mass and charge of the black holes that could be produced, and
physically it meant that the only black holes that could be pair
produced were those that are in thermodynamic equilibrium.
However, Wu \cite{WuSubMax}, and Bousso and Hawking
\cite{BoussoHawkSubMax} have shown that Euclidean solutions with
conical singularities (sub-maximal instantons) may also be used as
saddle points for the pair creation process, as long as the
spacelike boundary of the manifold is chosen in order to contain
the conical singularity and the metric is specified there. In this
way, pair creation of black holes whose horizons are not in
thermodynamic equilibrium is also allowed.

\subsection{\label{sec:Int-2}Historical overview on pair creation process in an external field}

We will describe the studies that have been done on pair creation
of black holes in an external field.

(i) The suggestion that the pair creation of black holes could
occur in quantized Einstein-Maxwell theory has been given by
Gibbons \cite{Gibbons-book} in 1986, who has proposed that
extremal black holes could be produced in a background magnetic
field and that the appropriate instanton describing the process
could be obtained by euclideanizing the extremal Ernst solution.
This idea has been recovered by Giddings and Garfinkle
\cite{GarfGidd} that confirmed the expectation of
\cite{Gibbons-book} and, in addition, they have constructed an
Ernst instanton that describes pair creation of nonextreme black
holes. The explicit calculation of the rate for this last process
has lead Garfinkle, Giddings and Strominger
\cite{GarfGiddStrom_Sbh} to conclude that the pair creation rate
of nonextreme black holes is enhanced relative to the pair
creation of monopoles and extreme black holes by a factor
$e^{S_{\rm bh}}$, where $S_{\rm bh}={\cal A}_{\rm bh}/4$ is the
Hawking-Bekenstein entropy of the black hole and ${\cal A}_{\rm
bh}$ is the area of the black hole event horizon. This issue of
black hole pair creation in a background magnetic field and the
above relation between the pair creation rate and the entropy has
been further investigated by Dowker, Gauntlett, Kastor and
Traschen \cite{DGKT}, by Dowker, Gauntlett, Giddings and Horowitz
\cite{DGGH}, and by Ross \cite{RossU(1)}, but now in the context
of the low energy limit of string theory and in the context of
five-dimensional Kaluza-Klein theory. To achieve their aim they
have worked with an effective dilaton theory which, for particular
values of the dilaton parameter, reduces to the above theories,
and they have explicitly constructed the dilaton Ernst instantons
that describe the process. The one-loop contribution to the
magnetic black hole pair creation problem has been given by Yi
\cite{YiPConeLoop}. Brown \cite{Brown,Brown2} has analyzed the
pair creation of charged black holes in an electric external
field. Hawking, Horowitz and Ross \cite{HawHorRoss} (see also
Hawking and Horowitz \cite{HawkHor}) have related the rate of pair
creation of extreme black holes with the area of the acceleration
horizon. In the nonextreme case, the rate has an additional
contribution from the area of the black hole horizon. From these
relations emerges an explanation for the fact, mentioned above,
that the pair creation rate of nonextreme black holes is enhanced
relative to the pair creation of extremal black holes by precisely
the factor $e^{{\cal A}_{\rm bh}/4}$. For a detailed discussion
concerning the reason why this factor involves only ${\cal A}_{\rm
bh}$ and not two times this value see also Emparan \cite{Emparan}.
It has to do with the fact that the internal microstates of two
members of the black hole pair are correlated.

(ii) The study of pair creation of de Sitter (dS) black holes has
been also investigated. Notice that the dS black hole solution can
be interpreted as a pair of dS black holes that are being
accelerated apart by the positive cosmological constant. The
cosmological horizon can be seen as an acceleration horizon that
impedes the causal contact between the two black holes, and this
analogy is perfectly identified for example when we compare the
Carter-Penrose diagrams of the C-metric and of the dS
Schwarzschild black hole, for example. The study on pair creation
of black holes in a dS background has begun in 1989 by Mellor and
Moss \cite{MelMos}, who have identified the gravitational
instantons that describe the process (see also Romans \cite{Rom}
for a detailed  construction of these instantons). The explicit
evaluation of the pair creation rates of neutral and charged black
holes accelerated by a cosmological constant has been done by Mann
and Ross \cite{MannRoss}. This process has also been discussed in
the context of the
 inflationary era undergone by the universe by Bousso and Hawking
 \cite{BoussoHawk}. Garattini \cite{GaratinniOneLoop},
 and  Volkov and Wipf \cite{VolkovWipf} have
 computed the one-loop factor for this pair creation process,
 something that in gravity quantum level is not an easy task.
Booth and Mann \cite{BooMann} have analyzed the cosmological pair
production of charged and rotating black holes. Pair creation of
dilaton black holes in a dS background has also been discussed by
Bousso \cite{BoussoDil}.

(iii) In 1995, Hawking and Ross \cite{HawkRoss-string} and
Eardley, Horowitz, Kastor and Traschen \cite{DougHorKastTras} have
discussed a process in which a cosmic string breaks and a pair of
black holes is produced at the ends of the string. The string
tension then pulls the black holes away, and the C-metric provides
the appropriate instantons to describe their creation. In order to
ensure that this process is physically consistent Ach\'ucarro,
Gregory and Kuijken \cite{AchGregKui}, and Gregory and Hindmarsh
\cite{GregHind} have shown that a conical singularity can be
replaced by a Nielson-Olesen vortix. This vortix can then pierce a
black hole \cite{AchGregKui}, or end at it \cite{GregHind}.
Moreover, it has been suggested that even topologically stable
strings can end at a black hole
\cite{HawkRoss-string}-\cite{PreskVil}.

(iv) We can also consider a pair creation process, analyzed by
Emparan \cite{Empar-string}, involving cosmic string breaking in a
background magnetic field. In this case the Lorentz force is in
excess or in deficit relative to the net force necessary to
furnish the right acceleration to the black holes, and this
discrepancy is eliminated by the string tension. The instantons
describing this process are a combination of the Ernst and
C-metric intantons.

(v) The gravitational repulsive energy of a domain wall provides
another mechanism for black hole pair creation. This process has
been analyzed by Caldwell, Chamblin and Gibbons
\cite{CaldChamGibb}, and by Bousso and Chamblin \cite{BouCham} in
a flat background, while in an anti-de Sitter background the pair
creation of topological  black holes (with hyperbolic topology)
has been analyzed by Mann \cite{MannAdS}.

Other studies concerning the process of pair creation in a
generalized background is done in \cite{Other}.

\subsection{\label{sec:Int-3}Pair creation of magnetic $\bm vs$ electric black holes}

It has been noticed that oddly the pair creation of electric black
holes was apparently enhanced relative to the pair creation of
magnetic black holes. This was a consequence of the fact that the
Maxwell action has opposite signs in the two cases. Now, this
discrepancy between the two pair creation rates was not consistent
with the idea that electric and magnetic black holes should have
identical quantum properties. This issue has been properly and
definitively clarified by Hawking and Ross \cite{HawkRoss} and by
Brown \cite{Brown2}, who have shown that the magnetic and electric
solutions differ not only in their actions, but also in the nature
of the boundaries conditions that can be imposed on them. More
precisely, one can impose the magnetic charge as a boundary
condition at infinity but, in the electric case, one instead
imposes the chemical potential as a boundary condition. As a
consequence they proposed that the electric action should contain
an extra Maxwell boundary term. This term cancels the opposite
signs of the Maxwell action, and the pair creation rate of
magnetic and electric black holes is equal.

\subsection{\label{sec:Int-4}Pair creation of black holes and the information loss problem}

The process of black hole pair creation gives also useful clues to
the discussion of the black hole information loss problem
\cite{InformationLossLOSS}. Due to the thermal Hawking radiation
the black holes evaporate. This process implies that one of the
following three scenarios occurs (see \cite
{InformationLossREVIEW} for reviews): (i) the information
previously swallowed to the black hole is destroyed, (ii) this
information is recovered to the exterior through the Hawking
radiation, or (iii) the endpoint of the evaporation is a Plank
scale remnant which stores the information. There are serious
difficulties associated to each one of this scenarios. Scenario
(i) implies non-unitarity and violation of energy conservation,
scenario (ii) implies violation of locality and causality, and the
main problem with scenario (iii) is that a huge energy is needed
in order to store all the information that has been swallowed by
the black hole, and a Planck scale remnant has very little energy.
Pair creation of black holes has been used to test these
scenarios. Indeed, it has been argued \cite{InformationLossREVIEW}
that if one demands preservation of unitarity and of locality then
a careful analysis of the one-loop contribution to the pair
creation process indicates that the Hawking process would leave
behind a catastrophic  infinite number of remnants. So the remnant
hypothesis seems to be discarded, although some escape solutions
can be launched \cite{InformationLossREVIEW}. On the other side,
Hawking, Horowitz and Ross \cite{HawHorRoss} have called attention
to the fact that the same instantons that describe pair creation
can, when reversed in time, describe their pair annihilation, as
long as the black holes have appropriate initial conditions such
that they come to rest at the right critical separation (this
annihilation process was also discussed by Emparan
\cite{Emparan}). One can then construct \cite{HawHorRoss} an
argument that  favors the information loss scenario: black holes
previously produced as a particle-antiparticle pair can accrete
information and annihilate, with their energy being given off as
electromagnetic and gravitational radiation. Therefore,  the
information loss scenario seems to occur at least in this
annihilation process.

\subsection{\label{sec:Int-5}Energy released during and after pair creation}
An important process that accompanies the production of the black
hole pair and the subsequent acceleration that they suffer is the
emission of electromagnetic and gravitational radiation. In an
asymptotically flat background, an estimate for the amount of
gravitational radiation radiated during the pair creation period
has been given by Cardoso, Dias and Lemos \cite{VitOscLem}:
$\Delta E =\frac{4 G c}{\pi} \frac{\gamma^3m^3}{\hbar}$, where $m$
is the mass of each one of the created black holes and
$\gamma=(1-v^2/c^2)^{-1/2}$ is the Lorentz factor. This value can
lead, under appropriate numbers of $m$ and $\gamma$ to huge
quantities, and is a very good candidate to emission of
gravitational radiation. For example, for black holes with 30
times the Planck mass and with $10\%$ of the velocity of light,
the gravitational energy released is $\Delta E\sim 10^{13}\,{\rm
J}$, which is about 100 times the rest energy of the pair.

The gravitational radiative properties of the resulting
accelerated black holes has been analyzed by Bi\v c\'ak, and
Pravda and Pravdova \cite{BPP}. In a dS background, the
gravitational radiation emitted by uniformly accelerated sources
without horizons has been analyzed by Bi\v c\'ak and Krtou\v s
\cite{BicKrt}, and the radiative properties of accelerated black
holes have been studied by Krtou\v s and Podolsk\' y
\cite{KrtPod}.

\subsection{\label{sec:Int-6}Plan of the paper}

In this paper we discuss the process in which a cosmic string
nucleates in a de Sitter (dS) background, and then breaks
producing a pair of black holes at its ends. Therefore, the energy
to materialize and accelerate the pair comes from the positive
cosmological constant and, in addition, from the string tension.
This process is a combination of the processes considered in (ii)
\cite{MelMos}-\cite{VolkovWipf} and in (iii)
\cite{HawkRoss-string}-\cite{GregHind}. The instantons for this
process can be constructed by analytically continuing  the dS
C-metric found by Pleba\'nski and Demia\'nski \cite{PlebDem} and
analyzed by Podolsk\'y and Griffiths \cite{PodGrif2}, and in
detail by Dias and Lemos \cite{OscLem_dS-C}.

The plan of this paper is as follows. In Sec.
\ref{sec:instanton-method}, we describe the semiclassical
instanton method used to evaluate the pair creation rate. In
section \ref{sec:dS C-inst} we construct, from the dS C-metric,
the instantons that describe the pair creation process. Then, in
section \ref{sec:Calc-I}, we explicitly evaluate the pair creation
rate for each one of the cases discussed in Sec.  \ref{sec:dS
C-inst}. In Sec.  \ref{sec:Entropy} we verify that the usual
relation between pair creation rate, entropy and total area holds
also for the pair creation process discussed in this paper.
Finally, in Sec.  \ref{sec:Conc} concluding remarks are presented.
In the Appendix a heuristic derivation of the pair creation rates
is given. Throughout this paper we use units in which
$G=c=\hbar=1$.

\section{\label{sec:instanton-method}Black hole pair creation
rate: the instanton method}
The pair creation of black holes in a de Sitter (dS) background is
described, according to the no-boundary proposal of Hartle and
Hawking \cite{HartleHawk}, by the propagation from nothing to a
3-surface boundary $\Sigma$. The amplitude for this process is
given by the wave function
 \begin{eqnarray}
\Psi(h_{ij},A_{i})=\int
d[g_{\mu\nu}]d[A_{\mu}]e^{-I(g_{\mu\nu},A_{\mu})}\:,
 \label{wave}
 \end{eqnarray}
where $h_{ij}$ and $A_{i}$ are the induced metric and
electromagnetic potential on the boundary
 $\Sigma=\partial {\cal M}$ of a compact manifold ${\cal M}$,
 $d[g_{\mu\nu}]$ is a measure on the space of the metrics $g_{\mu\nu}$
and $d[A_{\mu}]$ is a measure on the space of the Maxwell field
$A_{\mu}$, and $I(g_{\mu\nu},A_{\mu})$ is their Euclidean action.
The path integral is over all compact metrics and potentials on
manifolds ${\cal M}$ with boundary $\Sigma$, which agree with the
boundary data on $\Sigma$. For a detailed discussion of the
no-boundary proposal applied to the study of black hole pair
creation see Bousso and Chamblin \cite{BouCham}.

In the semiclassical instanton approximation, the dominant
contribution to the path integral comes from metrics and Maxwell
fields which are near the solutions (instantons) that extremalize
the Euclidean action and satisfy the boundary conditions. Thus,
considering small fluctuations around this solution, $g_{\mu\nu}
\rightarrow g_{\mu\nu}+\tilde{g}_{\mu\nu}$ and $A_{\mu}
\rightarrow A_{\mu}+\tilde{A}_{\mu}$, the action expands as
\begin{eqnarray}
I=I_{\rm inst}(g_{\mu\nu},A_{\mu})+
 \delta^2 I(\tilde{g}_{\mu\nu})+ \delta^2 I(\tilde{A}_{\mu})+\cdots\:,
 \label{Ifluct}
 \end{eqnarray}
where $\delta^2 I$ are quadratic in $\tilde{g}_{\mu\nu}$ and
$\tilde{A}_{\mu}$, and dots denote higher order terms. The wave
function, that describes the creation of a black hole pair from
nothing, is then given by $\Psi_{\rm inst}= B e^{-I_{\rm inst}}$,
where $I_{\rm inst}$ is the classical action of the gravitational
instanton that mediates the pair creation of black holes, and the
prefactor $B$ is the one loop contribution from the quantum
quadratic fluctuations in the fields, $\delta^2 I$. Similarly, the
wave function that describes the nucleation of a dS space with a
string from nothing is $\Psi_{\rm string} \propto e^{-I_{\rm
string}}$, and the wave function describing the nucleation of a dS
space from nothing is $\Psi_{\rm dS} \propto e^{-I_{\rm dS}}$.
 The nucleation probability of the dS
space from nothing, of the dS space with a string from nothing,
and of a space with a pair of black holes from nothing is then
given by $|\Psi_{\rm dS}|^2$, $|\Psi_{\rm string}|^2$ and
$|\Psi_{\rm inst}|^2$, respectively.

We may now ask four questions: what is the probability for (i)
pair creation of black holes in a dS spacetime, (ii) the
nucleation of a string in a dS background, (iii) the process in
which a string in a dS background breaks and a pair of black holes
is created, and (iv) the combined process (ii)+(iii). In the
process (i) the energy to materialize the pair comes only from the
positive cosmological constant background, $\Lambda$. The system
does not contain a string and the probability for this process has
been found in \cite{MannRoss}. The aim of the present paper is to
compute explicitly the probability for processes (ii)-(iv). It is
important to note that in the process (iii), one assumes that the
initial background contains a string, i.e. , the question that is
being asked is: given that the string is already present in our
initial system, what is the probability that it breaks and a pair
of black holes is produced and accelerated apart by $\Lambda$ and
by the string tension? On the other side, in (iv) one is asking:
starting from a pure dS background, what is the probability that a
string nucleates on it and then breaks forming a pair of black
holes? Naturally, the probability for process (iv) is  the product
of the probability for process (ii) and the probability for
process (iii).

According to the no-boundary proposal, the nucleation rate of a
string in a dS background is proportional to $|\Psi_{\rm
string}|^2 /|\Psi_{\rm dS}|^2$, i.e.,
\begin{eqnarray}
\Gamma_{\rm string/dS} \simeq \eta\bar{\eta} \, e^{- 2I_{\rm
string}+2I_{\rm dS} } \:.
 \label{PC-rate-string}
 \end{eqnarray}
The pair creation rate of black holes when a string breaks in a dS
background is given by
\begin{eqnarray}
\Gamma_{\rm BHs/string} \simeq \eta \, e^{- 2I_{\rm inst}+2I_{\rm
string} } \:,
 \label{PC-rate-breakstring}
 \end{eqnarray}
and the pair creation rate of black holes when process (iv) occurs
is given by the product of (\ref{PC-rate-string}) and
(\ref{PC-rate-breakstring}), i.e.,
\begin{eqnarray}
\Gamma_{\rm BHs/dS} \simeq \tilde{\eta} \, e^{- 2I_{\rm
inst}+2I_{\rm dS} } \:.
 \label{PC-rate}
 \end{eqnarray}
In Sec.  \ref{sec:Calc-I} we will find $I_{\rm inst}$ and $I_{\rm
string}$. In the three relations above, $\bar{\eta}$, $\eta$ and
$\tilde{\eta}$ are one-loop prefactors which will not be
considered in this paper. The evaluation of this one-loop
prefactor has been done only in a small number of cases, namely
for the vacuum background by Gibbons, Hawking and Perry
\cite{GibbHawPer}, for the Schwarzschild instanton by Gross, Perry
and Yaffe \cite{GrossPerryYaffe}, for other asymptotically flat
instantons by Young \cite{Young}, for the dS background by Gibbons
and Perry \cite{GibbPer} and Christensen and Duff
\cite{ChristDuff}, for the dS-Schwarzschild instanton by Ginsparg
and Perry \cite{GinsPerry}, Young \cite{YoungdS}, Volkov and Wipf
\cite{VolkovWipf} and Garattinni \cite{GaratinniOneLoop}, and for
the Ernst instanton by Yi \cite{YiPConeLoop}.

At this point we must specify the Euclidean action needed to
compute the path integral (\ref{wave}). This issue was analyzed
and clarified in detail by Hawking and Ross \cite{HawkRoss} and by
Brown \cite{Brown2}. Now, due to its relevance for the present
paper, we briefly discuss the main results of
\cite{HawkRoss,Brown2}. One wants to use an action for which it is
natural to fix the boundary data on $\Sigma$ specified in
(\ref{wave}). That is, one wants to use an action whose variation
gives the Euclidean equations of motion when the variation fixes
these boundary data on $\Sigma$ \cite{BrownYork}. In the magnetic
case this Euclidean action is the Einstein-Maxwell action with a
positive cosmological constant $\Lambda$ given by
 \begin{eqnarray}
I&=&-\frac{1}{16\pi}\int_{\cal M} d^4x\sqrt{g} \left (
R-2\Lambda-F^{\mu\nu}F_{\mu\nu} \right ) \nonumber \\
 & &-\frac{1}{8\pi}\int_{\Sigma=\partial {\cal M}} d^3x\sqrt{h}\, K \:,
 \label{I}
 \end{eqnarray}
where $g$ is the determinant of the Euclidean metric, $h$ is the
determinant of the induced metric on the boundary $\Sigma$, $R$ is
the Ricci scalar, $K$ is the trace of
 the extrinsic curvature $K_{ij}$ of the boundary, and
 $F_{\mu\nu}=\partial_{\mu}A_{\nu}-\partial_{\nu}A_{\mu}$ is
the Maxwell field strength of the gauge field $A_{\nu}$. Variation
of (\ref{I}) yields $\delta I=(\cdots )+
 \frac{1}{4\pi}\int_{\Sigma}
d^3x\sqrt{h}\, F^{\mu\nu}n_{\mu}\delta A_{\nu}$, where $(\cdots)$
represents terms giving the equations of motion plus gravitational
boundary terms that are discussed in \cite{BrownYork}, and
$n_{\mu}$ is the unit outward normal to $\Sigma$. Thus, variation
of (\ref{I}) gives the equations of motion as long as it is at
fixed gauge potential $A_i$ on the boundary. Now, for magnetic
black hole solutions, fixing the potential fixes the charge on
each of the black holes, since the magnetic charge is just given
by the integral of $F_{ij}$ over a 2-sphere lying in the boundary.
However, in the electric case, fixing $A_i$ can be regarded as
fixing a chemical potential $\omega$ which is conjugate to the
charge \cite{HawkRoss}. Holding the electric charge fixed is
equivalent to fixing $n_{\mu}F^{\mu i}$ on $\Sigma$, as the
electric charge is given by the integral of the dual of $F$ over a
2-sphere lying in $\Sigma$. Therefore in the electric case the
appropriate Euclidean action is \cite{HawkRoss}
 \begin{eqnarray}
I_{\rm el}=I-\frac{1}{4\pi}\int_{\Sigma=\partial {\cal M}}
d^3x\sqrt{h}\, F^{\mu\nu}n_{\mu}A_{\nu}\:,
 \label{I-electric}
 \end{eqnarray}
 where $I$ is defined in (\ref{I}).
Variation of action (\ref{I-electric}) yields $\delta I_{\rm
el}=(\cdots)+
 \frac{1}{4\pi}\int_{\Sigma}
d^3x \delta (\sqrt{h}\, F^{\mu\nu}n_{\mu}) A_{\nu}$, and thus it
gives the equations of motion when $\sqrt{h}\,n_{\mu} F^{\mu i}$,
and so the electric charge, is held fixed. Since $\int_{\cal M}
d^4x\sqrt{g} F^{\mu\nu}F_{\mu\nu}$ has opposite signs for dual
magnetic and electric solutions, if we took (\ref{I}) to evaluate
both the magnetic and electric actions we would conclude that the
pair creation of electric black holes would be enhanced relative
to the pair creation of magnetic black holes. This physically
unexpected result does not occur when one considers the
appropriate boundary conditions and includes the extra Maxwell
boundary term in (\ref{I-electric}).

We have to be careful \cite{GibbHawk,HawkRoss} when computing the
extra Maxwell boundary term in the electric action
(\ref{I-electric}). Indeed, we have to find a vector potential,
$A_{\nu}$, that is regular everywhere in the instanton, including
at the horizons. Usually, as we shall see, this requirement leads
to unusual choices for $A_{\nu}$. The need of this requirement is
easily understood if we take the example of the electric
Reissner-Nordstr\"{o}m solution \cite{GibbHawk,HawkRoss}. In this
case, normally, the gauge potential in Schwarzschild coordinates
is taken to be $A=-\frac{q}{r} \,dt$. However, this potential is
not regular at the horizon $r=r_+$, since $dt$ diverges there. An
appropriate choice that yields a regular electromagnetic potential
everywhere, including at the horizon is $A=-q (\frac{1}{r}-
\frac{1}{r_+}) \,dt$ or, alternatively, $A=-\frac{q}{r^2}\,t\,dr$.
To all these potentials corresponds the field strength
$F=-\frac{q}{r^2} \,dt\wedge dr$.

\section{\label{sec:dS C-inst} The \lowercase{d}S C-metric instantons}

The dS C-metric has been found by Pleba\'nski and Demia\'nski
\cite{PlebDem}. The physical properties and interpretation of this
solution have been analyzed by  Podolsk\'y and Griffiths
\cite{PodGrif2}, and in detail by Dias and Lemos
\cite{OscLem_dS-C}. The dS C-metric describes a pair of uniformly
accelerated black holes in a dS background, with the acceleration
being provided by the cosmological constant and, in addition, by a
string that connects the two black holes along their south poles
and pulls them away. The presence of the string is associated to
the conical singularity that exists in the south pole of the dS
C-metric (see, e.g., \cite{OscLem_dS-C,OscLem_AdS-C}). For a
detailed discussion on the properties of the dS C-metric we ask
the reader to see  \cite{OscLem_dS-C}. Here we will only mention
those which are really essential.

Following Sec.  \ref{sec:instanton-method}, in order to evaluate
the black hole pair creation rate we need to find the instantons
of the theory, i.e., we must look into the Euclidean section of
the dS C-metric and choose only those Euclidean solutions which
are regular in a way that will be explained soon. To obtain the
Euclidean section of the dS C-metric from the Lorentzian dS
C-metric we simply introduce an imaginary time coordinate
$\tau=-it$. Then the gravitational field of the Euclidean dS
C-metric is given by (see, e.g., \cite{OscLem_dS-C})
\begin{equation}
 d s^2 = [A(x+y)]^{-2} ({\cal F}d\tau^2+
 {\cal F}^{-1}dy^2+{\cal G}^{-1}dx^2+
 {\cal G}d\phi^2)\:,
 \label{C-metric PCdS}
 \end{equation}
\begin{eqnarray}
 & &{\cal F}(y) = -\frac{\Lambda+3A^2}{3A^2}
                     +y^2-2mAy^3+q^2A^2y^4, \nonumber \\
 & &{\cal G}(x) = 1-x^2-2mAx^3-q^2 A^2 x^4\:,
 \label{FG}
 \end{eqnarray}
where $\Lambda>0$ is the cosmological constant, $A>0$ is the
acceleration of the black holes, and $m$ and $q$ are the ADM mass
and electromagnetic charge of the non-accelerated black hole,
respectively. The Maxwell field in the magnetic case is given by
\begin{eqnarray}
 F_{\rm mag}=-q\, dx\wedge d\phi \:,
\label{F-mag}
\end{eqnarray}
while in the electric case it is given by
 \begin{eqnarray}
F_{\rm el}=-i\,q\, d\tau\wedge dy \:.
 \label{F-el}
\end{eqnarray}
The solution has a curvature singularity at $y=+\infty$ where the
matter source is.  The point $y=-x$ corresponds to a point that is
infinitely far away from the curvature singularity, thus as $y$
increases we approach the curvature singularity and $y+x$ is the
inverse of a radial coordinate. At most, ${\cal F}(y)$ can have
four real zeros which we label in ascending order by $y_{\rm
neg}<0<y_A\leq y_+ \leq y_-$. The roots $y_-$ and $y_+$ are
respectively the inner and outer charged black hole horizons, and
$y_A$ is an acceleration horizon which coincides with the
cosmological horizon and has a non-spherical shape. The negative
root $y_{\rm neg}$ satisfies $y_{\rm neg}<-x$ and thus has no
physical significance. The angular coordinate $x$ belongs to the
range $[x_\mathrm{s},x_\mathrm{n}]$ for which ${\cal G}(x)\geq 0$
(when we set $A=0$ we have $x_\mathrm{s}=-1$ and
$x_\mathrm{n}=+1$). In order to avoid a conical singularity in the
north pole, the period of $\phi$ must be given by
\begin{equation}
\Delta \phi=\frac{4 \pi}{|{\cal G}'(x_\mathrm{n})|}\:,
 \label{Period phi}
 \end{equation}
and this leaves a conical singularity in the south pole with
deficit angle
\begin{eqnarray}
\delta =2\pi \left ( 1-\frac{{\cal G}'(x_\mathrm{s})}{|{\cal
G}'(x_\mathrm{n})|} \right )\,.
 \label{conic-sing-dS}
 \end{eqnarray}
that signals the presence of a string with mass density $\mu
=\delta/(8\pi)$, and with pressure $p=-\mu<0$. When we set the
acceleration parameter $A$ equal to zero, the dS C-metric reduces
to the usual dS$-$Reissner-Nordstr\"{o}m or dS-Schwarzschild
solutions without conical singularities.

So far, we have described the solution that represents a pair of
black holes accelerated by the cosmological constant and by the
string tension. This solution describes the evolution of the black
hole pair after its creation. Now, we want to find a solution that
represents a string in a dS background. This solution will
describe the initial system, before the breaking of the cosmic
string that leads to the formation of the black hole pair. In
order to achieve our aim we note that at spatial infinity the
gravitational field of the Euclidean dS C-metric reduces to
\begin{eqnarray}
\!\!\!\!\!\!\!\!\! ds^2 \!\! &=& \!\!
\frac{1}{[A_0(x+y)]^2}{\biggl [}
 \left ( -\frac{\Lambda}{3A_0^{\,2}}-1+y^2 \right )dt^2
 \nonumber\\
 & & +  \frac{dy^2}{-\frac{\Lambda}{3A_0^{\,2}}-1+y^2}
+\frac{dx^2}{1-x^2}+ (1-x^2)d\phi_0^{\,2} {\biggr ]},
 \label{backg metric-dS}
\end{eqnarray}
and the Maxwell field goes to zero. $A_0$ is a constant that
represents a freedom in the choice of coordinates, and $-1 \leq x
\leq 1$. We want that this metric also describes the solution
before the creation of the black hole pair, i.e., we demand that
it describes a string with its conical deficit in a dS background.
Now, if we want to maintain the intrinsic properties of the string
during the process we must impose that its mass density and thus
its conical deficit remains constant. After the pair creation we
already know that the conical deficit is given by
(\ref{conic-sing-dS}). Hence, the requirement that the background
solution describes a dS spacetime with a conical deficit angle
given exactly by (\ref{conic-sing-dS}) leads us to impose that in
(\ref{backg metric-dS}) one has
\begin{eqnarray}
\Delta \phi_0=2\pi-\delta=2\pi \frac{{\cal
G}'(x_\mathrm{s})}{|{\cal G}'(x_\mathrm{n})|} \,.
 \label{delta phi0-dS}
 \end{eqnarray}
The arbitrary parameter $A_0$ can be fixed by imposing a matching
between (\ref{C-metric PCdS}) and (\ref{backg metric-dS}) at large
spatial distances \cite{HawHorRoss,HawkRoss-string}, yielding
$A_0^{\,2}=-A^2 [{\cal G}'(x_\mathrm{s})]^2/\left [ 2{\cal
G}''(x_\mathrm{s}) \right ]$.

Returning back to the euclidean dS C-metric (\ref{C-metric PCdS}),
in order to have a positive definite Euclidean metric we must
require that $y$ belongs to $y_A \leq y \leq y_+$. In general,
when $y_+ \neq y_-$, one has conical singularities at the horizons
$y=y_A$ and $y=y_+$. In order to obtain a regular solution we have
to eliminate the conical singularities at both horizons. This is
achieved by imposing that the period of $\tau$ is the same for the
two horizons, and is equivalent to requiring that the Hawking
temperature of the two horizons be equal. To eliminate the conical
singularity at $y=y_A$ the period of $\tau$ must be $\beta=2 \pi/
k_A$ (where $k_A$ is the surface gravity of the acceleration
horizon),
\begin{equation}
\beta=\frac{4 \pi}{|{\cal F}'(y_A)|}\:.
 \label{Period tau-yA PCdS}
 \end{equation}
 This choice for the period of $\tau$ also eliminates
simultaneously the conical singularity at the outer black hole
horizon, $y_+$, if and only if the parameters of the solution are
such that  the surface gravities of the  black hole and
acceleration horizons are equal ($k_+=k_A$), i.e.
\begin{equation}
 {\cal F}'(y_+)=-{\cal F}'(y_A)\:.
 \label{k+=kA PCdS}
 \end{equation}
There are two ways to satisfy this condition. One is a regular
Euclidean solution with $y_A \neq y_+$, and will be called
lukewarm C instanton. This solution requires the presence of an
electromagnetic charge. The other way is to have $y_A=y_+$, and
will be called Nariai C instanton. This last solution exists with
or without charge. When we want to distinguish them, they will be
labelled by charged Nariai and neutral Nariai C instantons,
respectively.

We now turn our attention to the case $y_+ = y_-$ and $y_A\neq
y_+$, which obviously requires the presence of charge. When this
happens the allowed range of $y$ in the Euclidean sector is simply
$y_A \leq y < y_+$. This occurs because when $y_+ = y_-$ the
proper distance along spatial directions between $y_A$ and $y_+$
goes to infinity. The point $y_+$ disappears from the $\tau, y$
section which is no longer compact but becomes topologically $S^1
\times {\mathbb{R}}$. Thus, in this case we have a conical
singularity only at $y_A$, and so we obtain a regular Euclidean
solution by simply requiring that the period of $\tau$ be equal to
(\ref{Period tau-yA PCdS}). We will label this solution by cold C
instanton. Finally, we have a special solution that satisfies
 $y_A=y_+=y_-$ and that is regular when condition
(\ref{Period tau-yA PCdS}) is satisfied. This instanton will be
called ultracold C instanton and can be viewed as a limiting case
of both the charged Nariai C instanton and cold C instanton.

Below, we will describe in detail each one of these four C
instantons, following the order: (A) lukewarm C instanton, (B)
cold C instanton, (C) Nariai C instanton, and (D) ultracold C
instanton. These instantons are the C-metric counterparts ($A\neq
0$) of the $A=0$ instantons that have been constructed from the
Euclidean section of the dS$-$Reissner-Nordstr\"{o}m solution
($A=0$) \cite{MelMos,Rom,MannRoss,BooMann}. The original name of
the $A=0$ instantons is associated to the relation between their
temperatures: $T_{\rm lukewarm}>T_{\rm cold}>T_{\rm
ultracold}>T_{\rm Nariai}=0$. This relation is preserved by their
C-metric counterparts discussed in this paper, and we preserve the
$A=0$ nomenclature. The ultracold instanton could also, very
appropriately, be called Nariai Bertotti-Robinson instanton (see
\cite{OscLem_nariai}). These four families of instantons will
allow us to calculate the pair creation rate of accelerated
dS$-$Reissner-Nordstr\"{o}m black holes in Sec. \ref{sec:Calc-I}.

As is clear from the above discussion, when the charge vanishes
the only regular Euclidean solution that can be constructed is the
neutral Nariai C instanton. The same feature is present in the
$A=0$ case where only the neutral Nariai instanton is available
\cite{GinsPerry,MannRoss,BoussoHawk,VolkovWipf}.

\subsection{\label{sec:Lukewarm-inst}The lukewarm C instanton}

For the lukewarm C instanton the gravitational field is given by
(\ref{C-metric PCdS}) with the requirement that ${\cal F}(y)$
satisfies ${\cal F}(y_+)=0={\cal F}(y_A)$ and ${\cal
F}'(y_+)=-{\cal F}'(y_A)$. In this case we can then write (onwards
the subscript ``$\ell$" means lukewarm)
\begin{eqnarray}
{\cal F}_{\rm \ell}(y)&=&-\left ( \frac{y_A \: y_+}{y_A+y_+}
\right )^2 \left
( 1-\frac{y}{y_A} \right ) \left ( 1-\frac{y}{y_+} \right ) \nonumber \\
 & & \times \left ( 1+\frac{y_A+y_+}{y_A \:y_+}\,y-\frac{y^2}{y_A \:y_+} \right
 )\:,
 \label{F-luk}
 \end{eqnarray}
with
\begin{eqnarray}
y_A &=&  \frac{1-\alpha}{2mA}\,, \:\:\:\:\:\:\:\: y_+ = \frac{1+\alpha}{2mA}\,, \nonumber \\
 \alpha &=& \sqrt{1-\frac{4m}{\sqrt{3}}\sqrt{\Lambda+3A^2}} \:.
 \label{yA-luk}
 \end{eqnarray}
 The parameters
$A$, $\Lambda$, $m$ and $q$, written as a function of $y_A$ and
$y_+$, are
\begin{eqnarray}
& & \frac{\Lambda}{3A^2} =  \left ( \frac{y_A \: y_+}{y_A+y_+}
\right )^2\,, \nonumber \\
& & mA=(y_A+y_+)^{-1}=qA \,.
 \label{zeros-luk}
 \end{eqnarray}
Thus, the mass and the charge of the lukewarm C instanton are
necessarily equal, $m=q$, as occurs with its $A=0$ counterpart,
the lukewarm instanton \cite{MelMos,Rom,MannRoss,BooMann}. The
demand that $\alpha$ is real requires that
\begin{eqnarray}
  0< m_{\rm \ell}\leq \frac{\sqrt{3}}{4} \frac{1}{\sqrt{\Lambda+3A^2}}\:,
 \label{mq-luk}
 \end{eqnarray}
so the lukewarm C instanton has a lower maximum mass and a lower
maximum charge than the $A=0$ lukewarm instanton
\cite{MelMos,Rom,MannRoss,BooMann} and, for a fixed $\Lambda$, as
the acceleration parameter $A$ grows this maximum value decreases
monotonically. For a fixed $\Lambda$ and for a fixed mass below
$\sqrt{3/(16 \Lambda)}$, the maximum value of the acceleration is
$\sqrt{1/(4m)^2-\Lambda/3}$.

As we said, the allowed range of $y$ in the Euclidean sector is
$y_A \leq y \leq y_+$. Then, the period of $\tau$, (\ref{Period
tau-yA PCdS}), that avoids the conical singularity at both
horizons is
\begin{equation}
\beta_{\rm \ell}=\frac{8 \pi \,m A}{\alpha(1-\alpha^2)}\,,
 \label{beta-luk}
 \end{equation}
and $T_{\rm \ell}=1/\beta_{\rm \ell}$ is the common temperature of
the two horizons.

Using the fact that ${\cal G}(x)=-\Lambda/(3A^2)-{\cal F}(-x)$
[see (\ref{FG})] we can write
\begin{eqnarray}
{\cal G}_{\rm \ell}(x) = 1-x^2 \left ( 1+m A\, x \right )^2 \:,
 \label{G-luk}
 \end{eqnarray}
 and the only real zeros of ${\cal G}_{\rm \ell}(x)$ are the south and
 north pole
\begin{eqnarray}
& & \!\!\!\!\!\!\!\!\!\!\!\!\!\!\!\!\!\!  x_\mathrm{s} =
\frac{-1+\omega_-}{2mA}<0\,, \:\:\:\:\:\:\:\:
x_\mathrm{n}  = \frac{-1+\omega_+}{2mA}>0\,, \\
 & & \!\!\!\!\!\!\!\!\!\!\!\!\!\!\!\!\!\!
  {\rm with}\:\:\:\: \omega_{\pm} \equiv \sqrt{1\pm 4mA} \:.
 \nonumber
 \label{polos-luk}
 \end{eqnarray}
When $A$ goes to zero we have $x_\mathrm{s}\rightarrow -1$ and
$x_\mathrm{n}\rightarrow +1$.
 The period of $\phi$, (\ref{Period phi}), that avoids the
conical singularity at the north pole (and leaves one at the south
pole responsible for the presence of the string) is
\begin{equation}
\Delta \phi _{\rm \ell}=\frac{8 \pi\,m A}{\omega_+(\omega^2_+ -1)}
\leq 2 \pi\:.
 \label{Period phi-luk}
 \end{equation}
 When $A$ goes to zero we have $\Delta \phi _{\rm \ell} \rightarrow 2 \pi$
 and the conical singularity disappears.

The topology of the lukewarm C instanton is $S^2 \times S^2$
 ($0\leq \tau \leq \beta_{\rm \ell}$,
 $y_A \leq y \leq y_+$, $x_\mathrm{s}\leq x \leq x_\mathrm{n}$,
and $0 \leq \phi \leq \Delta \phi _{\rm \ell}$). The Lorentzian
sector describes two dS black holes being accelerated by the
cosmological background and by the string, so this instanton
describes pair creation of nonextreme black holes with $m=q$.

\subsection{\label{sec:Cold-inst}The cold C instanton}
The gravitational field of the cold C instanton  is given by
(\ref{C-metric PCdS}) with the requirement that the size of the
outer charged black hole horizon $y_+$ is equal to the size of the
inner charged horizon $y_-$. Let us label this degenerated horizon
by $\rho$: $y_+=y_-\equiv \rho$ and $\rho > y_A$.  In this case,
the function ${\cal F}(y)$ can be written as (onwards the
subscript ``${\rm c}$" means cold)
\begin{eqnarray}
{\cal F}_{\rm c}(y)=\frac{\rho^2-3\gamma}{\rho^4}
 (y-y_{\rm neg})(y-y_A)(y-\rho)^2\:,
 \label{F-cold}
 \end{eqnarray}
with
\begin{eqnarray}
\gamma=\frac{\Lambda+3A^2}{3A^2}\:,
 \label{gamma-PCdS}
 \end{eqnarray}
  and the roots $\rho$,
$y_{\rm neg}$ and $y_A$ are given by
\begin{eqnarray}
& & \rho =\frac{3m}{4q^2A}
 \left ( 1+ \sqrt{1-\frac{8}{9}\frac{q^2}{m^2}} \:\right )
 \:,   \label{zerosy1-cold} \\
& & y_{\rm neg} =\frac{\gamma \rho}{\rho^2-3\gamma}
 \left ( 1- \sqrt{\frac{\rho^2-2\gamma}{\gamma}} \:\right )
 \:,   \label{zerosy2-cold}\\
& & y_A =\frac{\gamma \rho}{\rho^2-3\gamma}
 \left ( 1+ \sqrt{\frac{\rho^2-2\gamma}{\gamma}} \:\right )
 \:.
 \label{zerosy3-cold}
 \end{eqnarray}
The mass and the charge parameters of the solution are written as
a function of $\rho$ as
\begin{eqnarray}
& & m =\frac{1}{A\rho}
 \left ( 1- \frac{2\gamma}{\rho^2} \right )
 \:,  \nonumber \\
& & q^2 =\frac{1}{A^2\rho^2}
 \left ( 1- \frac{3\gamma}{\rho^2} \right )
 \:,
 \label{mq PCdS}
 \end{eqnarray}
and, for a fixed $A$ and $\Lambda$, the ratio $q/m$ is higher than
$1$. The conditions $\rho > y_A$ and $q^2 > 0$ require that, for
the cold C instanton, the allowed range of $\rho$ is
\begin{eqnarray}
\rho>\sqrt{6\gamma} \:.
 \label{range-gamma-cold}
\end{eqnarray}
The value of $y_A$ decreases monotonically with $\rho$ and we have
$\sqrt{\gamma}<y_A<\sqrt{6\gamma}$.
 The mass and the charge of the cold C instanton are also monotonically
decreasing functions of $\rho$, and as we come from $\rho=+\infty$
into $\rho=\sqrt{6\gamma}$ we have
\begin{eqnarray}
& &  0< m_{\rm c}< \frac{\sqrt{2}}{3}
  \frac{1}{\sqrt{\Lambda+3A^2}}\:, \\
& & 0< q_{\rm c}< \frac{1}{2}
  \frac{1}{\sqrt{\Lambda+3A^2}}\:,
 \label{mq-cold}
 \end{eqnarray}
so the cold C instanton has a lower maximum mass and a lower
maximum charge than the $A=0$ cold instanton, and, for a fixed
$\Lambda$, as the acceleration parameter $A$ grows this maximum
value decreases monotonically. For a fixed $\Lambda$ and for a
fixed mass below $\sqrt{2/(9 \Lambda)}$, the maximum value of the
acceleration is $\sqrt{2/(27m^2)-\Lambda/3}$.

As we have already said, the allowed range of $y$ in the Euclidean
sector is $y_A \leq y < y_+$ and does not include $y=y_+$. Then,
the period of $\tau$, (\ref{Period tau-yA PCdS}), that avoids the
conical singularity at the only  horizon of the cold C instanton
is
\begin{equation}
\beta_{\rm c}=\frac{2 \pi
\rho^3}{(y_A-\rho)^2\sqrt{\gamma(\rho^2-2\gamma)}}\:,
 \label{beta-cold}
 \end{equation}
and $T_{\rm c}=1/\beta_{\rm c}$ is the temperature of the
acceleration horizon.

In what concerns the angular sector of the cold C instanton,
${\cal G}(x)$ is given by (\ref{FG}), and its only real zeros are
the south and north pole,
\begin{eqnarray}
x_\mathrm{s} &=&  -p + \frac{h}{2}-\frac{m}{2q^2A}<0\,, \nonumber \\
 x_\mathrm{n}  &=& p +
\frac{h}{2}-\frac{m}{2q^2A}>0 \:,
 \label{polos-cold}
 \end{eqnarray}
with
\begin{eqnarray}
p &=& \frac{1}{2} \left (  -\frac{s}{3}+\frac{2m^2}{q^4A^2}
  -\frac{1-12 q^2A^2}{3s q^4A^4} -\frac{4}{3q^2A^2}
  + n\right )^{1/2} \:,  \nonumber \\
n &=&  \frac{-m^3+mq^2}{2hq^6 A^3} \:,  \nonumber \\
h &=& \sqrt{\frac{s}{3}+\frac{m^2}{q^4A^2}+\frac{1-12 q^2A^2}{3s q^4A^4}-\frac{2}{3q^2A^2}} \:, \nonumber \\
 s &=& \frac{1}{2^{1/3} q^2A^2}\left (
\lambda-\sqrt{\lambda^2-4(1-12 q^2A^2)^3} \right )^{1/3} \:,
\nonumber \\
\lambda &=& 2 - 108 m^2A^2 + 72 q^2A^2\:,
  \label{acess-zeros-ang PCdS}
 \end{eqnarray}
where $m$ and $q$ are fixed by  (\ref{mq PCdS}), for a given $A$,
$\Lambda$ and $\rho$. When $A$ goes to zero we have
$x_\mathrm{s}\rightarrow -1$ and $x_\mathrm{n}\rightarrow +1$. The
period of $\phi$, $\Delta \phi _{\rm c}$, that avoids the conical
singularity at the north pole (and leaves one at the south pole
responsible for the presence of the string) is given by
(\ref{Period phi}) with $x_\mathrm{n}$ defined in
(\ref{polos-cold}).

The topology of the cold C instanton is ${\mathbb{R}}^2 \times
S^2$, since $y=y_+=\rho$ is at an infinite proper distance ($0
\leq \tau \leq \beta_{\rm c}$, $y_A \leq y < y_+$,
 $x_\mathrm{s}\leq x \leq x_\mathrm{n}$,
and $0 \leq \phi \leq \Delta \phi _{\rm c}$). The surface
$y=y_+=\rho$ is then an internal infinity boundary that will have
to be taken into account in the calculation of the action of the
cold C instanton (see Sec.  \ref{sec:Cold-rate}). The Lorentzian
sector of this cold case describes two extreme ($y_+=y_-$) dS
black holes being accelerated by the cosmological background and
by the string, and the cold C instanton describes pair creation of
these extreme black holes.

\subsection{\label{sec:Nariai-inst}The Nariai C instanton}
In the case of the Nariai C instanton, we  require that the size
of the acceleration horizon $y_A$ is equal to the size of the
outer charged horizon $y_+$. Let us label this degenerated horizon
by $\rho$: $y_A=y_+\equiv \rho$ and $\rho < y_-$.
 In this case, the function ${\cal F}(y)$ can be written as (onwards the subscript
 ``${\rm N}$" means Nariai)
\begin{eqnarray}
{\cal F}_{\rm N}(y) =\frac{\rho^2-3\gamma}{\rho^4}
 (y-y_{\rm neg})(y-y_-)(y-\rho)^2\:,
 \label{F-cNariai}
 \end{eqnarray}
where $\gamma$ is defined by (\ref{gamma-PCdS}), the roots $\rho$
and $y_{\rm neg}$ are given by (\ref{zerosy1-cold}) and
(\ref{zerosy2-cold}), and $y_-$ is given by $y_- =\frac{\gamma
\rho}{\rho^2-3\gamma}
 \left ( 1+ \sqrt{\frac{\rho^2-2\gamma}{\gamma}} \:\right )$.
The mass and the charge of the solution are defined as a function
of $\rho$ by (\ref{mq PCdS}). The conditions $\rho < y_-$ and $q^2
\geq 0$ require that for the Nariai C instanton, the allowed range
of $\rho$ is
\begin{eqnarray}
 \sqrt{3\gamma} \leq \rho<\sqrt{6\gamma}\:.
 \label{range-gamma-cNariai}
\end{eqnarray}
The value of $y_-$ decreases monotonically with $\rho$ and we have
$\sqrt{6\gamma}<y_-<+\infty$. Contrary to the cold C instanton,
the mass and the charge of the Nariai C instanton are
monotonically increasing functions of $\rho$, and as we go from
$\rho=\sqrt{3\gamma}$ to $\rho=\sqrt{6\gamma}$ we have
\begin{eqnarray}
& &  \frac{1}{3}
  \frac{1}{\sqrt{\Lambda+3A^2}}\leq m_{\rm N}< \frac{\sqrt{2}}{3}
  \frac{1}{\sqrt{\Lambda+3A^2}}\:, \nonumber \\
& & 0\leq q_{\rm N}< \frac{1}{2}
  \frac{1}{\sqrt{\Lambda+3A^2}}\:.
 \label{mq-cNariai PCdS}
 \end{eqnarray}
Note that $\rho= \sqrt{3\gamma}$ implies $q=0$.  For a fixed
$\Lambda$ and for a  mass fixed between $\sqrt{1/(9 \Lambda)}\leq
m<\sqrt{2/(9 \Lambda)}$, the acceleration varies as
$\sqrt{1/(27m^2)-\Lambda/3} \leq A <\sqrt{2/(27m^2)-\Lambda/3}$.

At this point, one has apparently a problem that is analogous to
the one that occurs with the $A=0$ neutral Nariai instanton
\cite{GinsPerry} and with the $A=0$ charged Nariai instanton
\cite{MannRoss,HawkRoss}. Indeed, as we said in the beginning of
this section, the allowed range of $y$ in the Euclidean sector is
$y_A \leq y \leq y_+$ in order to obtain a positive definite
metric. But in the Nariai case $y_A=y_+$, and so it seems that we
are left with no space to work with in the Euclidean sector.
However, as in \cite{GinsPerry,MannRoss,HawkRoss}, the proper
distance between $y_A$ and $y_+$ remains finite as $y_A
\rightarrow y_+$, as is shown in detail in \cite{OscLem_nariai}
where the Nariai C-metric is constructed and analyzed. In what
follows we briefly exhibit the construction. We first set
$y_A=\rho-\varepsilon$ and $y_+=\rho+\varepsilon$, in order that
$\varepsilon<<1$ measures the deviation from degeneracy, and the
limit $y_A\rightarrow y_+$ is obtained when $\varepsilon
\rightarrow 0$. Now, we introduce a new time coordinate
$\tilde{\tau}$, $\tau= \frac{1}{\varepsilon {\cal
K}}\,\tilde{\tau}$, and a new radial coordinate $\chi$,
$y=\rho+\varepsilon \cos\chi$, where $\chi=0$ and $\chi=\pi$
correspond, respectively, to the horizons $y_+$ and $y_A$, and
\begin{eqnarray}
{\cal K} = \frac{2(\Lambda+3A^2)}{A^2\rho^2}-1\:.
 \label{Kfactor PCdS}
\end{eqnarray}
Condition (\ref{range-gamma-cNariai})  implies $0<{\cal K}\leq 1$
with $q=0\Rightarrow {\cal K}=1$. Then, in the limit $\varepsilon
\rightarrow 0$, from  (\ref{C-metric PCdS}) and (\ref{F-cNariai}),
we obtain the gravitational field of the Nariai C instanton
\begin{eqnarray}
d s^2 &=& \frac{{\cal R}^2(x)}{{\cal K}}
\left (\sin^2\chi\, d\tilde{\tau}^2 +d\chi^2\right ) \nonumber \\
&+&
 {\cal R}^2(x)\left [{\cal G}^{-1}(x)dx^2+ {\cal G}(x)d\phi^2 \right ]
 \:.
 \label{Nariai-C-Metric PCdS}
\end{eqnarray}
where $\chi$ runs from $0$ to $\pi$, and
\begin{eqnarray}
 {\cal R}^2(x)&=& \left [ A(x+\rho) \right ]^{-2}\:.
\label{Rfactor PCdS}
\end{eqnarray}
In the new coordinates $\tilde{\tau}$ and $\chi$, the Maxwell
field for the magnetic case is still given by (\ref{F-mag}), while
in the electric case we have now
 \begin{eqnarray}
 F_{\rm el}=i\,\frac{q}{{\cal K}}\,\sin \chi \,d\tilde{\tau}\wedge d\chi\:.
\label{F-el-Nariai PCdS}
\end{eqnarray}
The period of $\tilde{\tau}$ of the Nariai C instanton is simply
$\beta_{\rm N}=2 \pi$.
The function ${\cal G}(x)$ is given by (\ref{FG}), with $m$ and
$q$ fixed by
 (\ref{mq PCdS}) for a given $A$, $\Lambda$ and $\sqrt{3\gamma} <
 \rho<\sqrt{6\gamma}$. Under these conditions the south and north
 pole (which are the only real roots) are also given by
 (\ref{polos-cold}) and (\ref{acess-zeros-ang PCdS}). The period of
$\phi$, $\Delta \phi _{\rm N}$, that avoids the conical
singularity at the north pole (and leaves one at the south pole
responsible for the presence of the string) is given by
(\ref{Period phi}) with $x_\mathrm{n}$ defined in
(\ref{polos-cold}).

The topology of the Nariai C instanton is $S^2 \times S^2$
 ($0 \leq \tilde{\tau} \leq \beta_{\rm N}$, $0\leq \chi \leq \pi$,
 $x_\mathrm{s}\leq x \leq x_\mathrm{n}$,
and $0 \leq \phi \leq \Delta \phi _{\rm N}$). The Nariai C
instanton transforms into the Nariai instanton when we take the
limit $A=0$ \cite{OscLem_nariai}.
 The Lorentzian sector is conformal to the direct topological product of
 $dS_2 \times S^2$, i.e., of a
(1+1)-dimensional de Sitter spacetime with a deformed 2-sphere of
fixed size. To each point in the sphere corresponds a $dS_2$
spacetime, except for one point - the south pole - which
corresponds a $dS_2$ spacetime with a string \cite{OscLem_nariai}.
When we set $A=0$ the $S^2$ is a round 2-sphere free of the
conical singularity and so, without the string. In this case it
has been shown \cite{GinsPerry,BoussoHawk} that the Nariai
solution decays through the quantum tunnelling process into a
slightly non-extreme dS black hole pair (for a complete review  on
this subject see, e.g., \cite{Bousso60y,OscLem_nariai}). We then
naturally expect that an analogous quantum instability is present
in the Nariai C-metric. Therefore, the Nariai C instanton
describes the creation of a Nariai C universe  that then decays
into a slightly non-extreme ($y_A \sim y_+$) pair of black holes
accelerated by the cosmological background and by the string.

We recall again that the neutral Nariai C instanton with
$m=\frac{1}{3} \frac{1}{\sqrt{\Lambda+3A^2}}$ is the only regular
Euclidean solution that can be constructed from the dS C-metric
when the charge vanishes. The same feature is present in the $A=0$
case where only the neutral Nariai instanton with $m=\frac{1}{3}
\frac{1}{\sqrt{\Lambda}}$ is available
\cite{GinsPerry,MannRoss,BoussoHawk,VolkovWipf}.
\subsection{\label{sec:Ultracold-inst}The ultracold C instanton}

In the case of the ultracold C instanton, we  require that  the
size of the three horizons ($y_A$, $y_+$ and $y_-$) is equal, and
let us label this degenerated horizon by $\rho$:
$y_A=y_+=y_-\equiv \rho$.
 In this case, the function ${\cal F}(y)$ can be written as (onwards
the subscript ``${\rm u}$" means ultracold)
\begin{eqnarray}
{\cal F}_{\rm u}(y) =\frac{\rho^2-3\gamma}{\rho^4}
 (y-y_{\rm neg})(y-\rho)^3\:,
 \label{F-ultra}
 \end{eqnarray}
where $\gamma$ is defined by (\ref{gamma-PCdS}) and the roots
$\rho$ and $y_{\rm neg}$ are given by (\ref{zerosy1-cold}) and
(\ref{zerosy2-cold}), respectively. Given the values of $A$ and
$\Lambda$, $\rho$ can take only the value
\begin{eqnarray}
\rho=\sqrt{6\gamma} \:.
 \label{range-gamma-ultra}
\end{eqnarray}
 The mass and the charge of the solution, defined by (\ref{mq-cold}), are then given by
\begin{eqnarray}
& &m_{\rm u} =\frac{\sqrt{2}}{3}
 \sqrt{\frac{1}{\Lambda+3A^2}}
 \:,  \nonumber \\
& & q_{\rm u} =\frac{1}{2}\sqrt{\frac{1}{\Lambda+3A^2}}
 \:,
 \label{mq-ultra}
 \end{eqnarray}
and these values are the maximum values of the mass and charge of
both the cold and charged Nariai instantons.

Being a limiting case of both the cold C instanton and of the
charged Nariai C instanton, the ultracold C instanton presents
similar features. The appropriate analysis of this solution (see
\cite{OscLem_nariai} for a detailed discussion) requires that we
first set $\rho=\sqrt{6\gamma}-\varepsilon$ and
$y_-=\sqrt{6\gamma}+\varepsilon$, with $\varepsilon<<1$. Then, we
introduce a new time coordinate $\tilde{\tau}$, $\tau= \frac{1}{2
\varepsilon^2 {\cal K}}\,\tilde{\tau}$, and a new radial
coordinate $\chi$, $y=\sqrt{6\gamma}+
 \varepsilon \cosh(\sqrt{2\varepsilon{\cal K}}\:\chi)$,
where ${\cal K} =\frac{1}{3}
 \sqrt{\frac{2A^2}{\Lambda+3A^2}}$.
Finally, in the limit $\varepsilon \rightarrow 0$, from
(\ref{C-metric PCdS}) and  (\ref{F-ultra}), we obtain the
gravitational field of the ultracold C instanton
\begin{eqnarray}
& & d s^2 = {\cal R}^2(x) \left [\chi^2\, d\tilde{\tau}^2 +d\chi^2
 +{\cal G}^{-1}(x)dx^2+ {\cal G}(x)d\phi^2 \right ],
 \nonumber \\
& & {\rm with} \qquad {\cal R}^2(x)=\left
(Ax+\sqrt{2(\Lambda+3A^2)}\right )^{-2}\:.
 \label{ultracold}
\end{eqnarray}
Notice that the spacetime factor $\chi^2\, d\tilde{\tau}^2
+d\chi^2$ is just Euclidean space in Rindler coordinates, and
therefore, under the usual coordinate transformation, it can be
putted in the form $dT^2+dX^2$. $\chi=0$ corresponds to the
Rindler horizon and $\chi=+\infty$ corresponds an internal
infinity boundary. In the new coordinates $\tilde{\tau}$ and
$\chi$,  the Maxwell field for the magnetic case is still given by
(\ref{F-mag}), while in the electric case we have now
 \begin{eqnarray}
 F_{\rm el}=-i\,q\,\chi \,d\tilde{\tau}\wedge d\chi\:.
\label{F-el-ultracold}
\end{eqnarray}
The period of $\tilde{\tau}$ of the ultracold C instanton is
simply $\beta_{\rm u}=2 \pi$.

 The function ${\cal G}(x)$ is given
by (\ref{FG}), with $m$ and $q$ fixed by
  (\ref{mq-ultra}) for a given $A$ and $\Lambda$. Under these
 conditions the south and north pole (which are the only real roots)
 are also given by  (\ref{polos-cold}) and (\ref{acess-zeros-ang PCdS}).
The period of $\phi$, $\Delta \phi _{\rm u}$, that avoids the
conical singularity at the north pole  is given by (\ref{Period
phi}) with $x_\mathrm{n}$ defined in (\ref{polos-cold}).

The topology of the ultracold C instanton is ${\mathbb{R}}^2
\times S^2$, since $\chi=+\infty$ is at an
 infinite proper distance
 ($0 \leq \tilde{\tau} \leq \beta_{\rm u}$, $0\leq \chi \leq \infty$,
 $x_\mathrm{s}\leq x \leq x_\mathrm{n}$,
 and $0 \leq \phi \leq \Delta \phi _{\rm u}$).
The surface $\chi=+\infty$ is then an internal infinity boundary
that will have to be taken into account in the calculation of the
action of the ultracold C instanton (see Sec.
\ref{sec:Ultracold-rate}). The ultracold C instanton transforms
into the ultracold instanton when we take the limit $A=0$
\cite{OscLem_nariai}.
 The Lorentzian sector is conformal to the direct topological product of
 ${\mathbb{M}}^{1,1}\times S^2$, i.e., of a
(1+1)-dimensional Minkowski spacetime with a deformed 2-sphere of
fixed size. To each point in the sphere corresponds a
${\mathbb{M}}^{1,1}$ spacetime, except for one point - the south
pole - which corresponds a ${\mathbb{M}}^{1,1}$ spacetime with a
string \cite{OscLem_nariai}. We can, appropriately, label this
solution as Nariai Bertotti-Robinson C universe (see
\cite{OscLem_nariai}). When we set $A=0$ the $S^2$ is a round
2-sphere free of the conical singularity and so, without the
string. In an analogous way to the Nariai C universe, the Nariai
Bertotti-Robinson C universe is unstable and decays into a
slightly non-extreme ($y_A \sim y_+ \sim y_-$) pair of black holes
accelerated by the cosmological background and by the string. The
ultracold C instanton mediates this decay.

\begin{figure}[t]
\includegraphics*[height=2.1in]{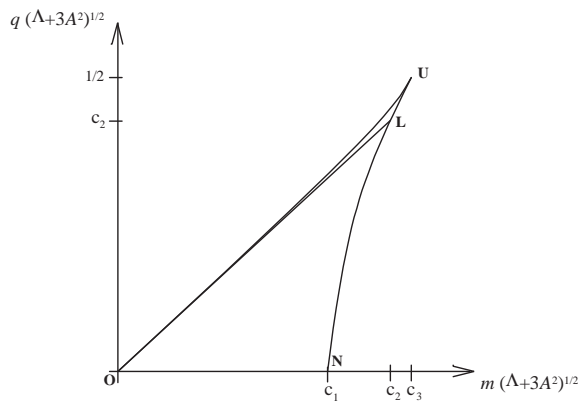}
\caption{\label{mq-fig}
 Relation $q\sqrt{\Lambda+3A^2}$ vs
$m\sqrt{\Lambda+3A^2}$ for a fixed value of $A$ and $\Lambda$ for
the four dS C instantons. $OL$ represents the lukewarm C instanton
($m=q$), $OU$ represents the cold C instanton, $NLU$ represents
the Nariai C instanton, and $U$ represents the ultracold C
instanton. Point $N$ represents the neutral Nariai C instanton,
the only instanton for the uncharged case. When we set $\Lambda=0$
we obtain the charge/mass relation of the flat C-metric and flat
Ernst instantons, and setting $A=0$ yields the charge/mass
relation for the dS instantons \cite{MannRoss}. This reveals the
close link that exists between the instantons that describe the
pair creation process in different backgrounds. $c_1=1/3$,
$c_2=\sqrt{3}/4$ and $c_3=\sqrt{2}/3$.
 }
\end{figure}

The allowed range of $m$ and $q$ for each one of the four C
instantons is sketched in Fig. \ref{mq-fig}.
\section{\label{sec:Calc-I}Calculation of the black hole pair creation rates}

In the last section we have found the regular dS C instantons that
can be interpreted as describing a circular motion in the
Euclidean sector of the solution with a period $\beta$. Each of
these instantons is the Euclidean continuation of a Lorentzian
solution that describes a pair of black holes that start at rest
at the creation moment, and then run hyperbolically in opposite
directions, due to a string that accelerates them. Now, in order
to compute the pair creation rate of the corresponding black
holes, we have to slice the instanton, the Euclidean trajectory,
in half along $\tau=0$ and $\tau=\beta/2$, where the velocities
vanish. The resulting geometry is precisely that of the moment of
closest approach of the black holes in the Lorentzian sector of
the dS C-metric, and in this way the Euclidean and Lorentzian
solutions smoothly match together. In particular, the extrinsic
curvature vanishes for both surfaces and therefore, they can be
glued to each other.

In this section, we will compute the black hole pair creation
rates given in (\ref{PC-rate-breakstring}) or (\ref{PC-rate}) for
each one of the four cases considered in Sec.  \ref{sec:dS
C-inst}. Moreover, we will also compute the pair creation rate of
black holes whose nucleation process is described by sub-maximal
instantons. For the reason explained in Sec.
\ref{sec:instanton-method}, the magnetic action will be evaluated
using (\ref{I}), while the electric action will be computed using
(\ref{I-electric}).  In general, in (\ref{I}) and
(\ref{I-electric}) we identify $\Sigma=\partial {\cal{M}}$ with
the surfaces of zero extrinsic curvature discussed just above. It
then follows that the boundary term $\int_{\Sigma} d^3x\sqrt{h}\,
K$ vanishes. However, as we already mentioned in Sec. s
\ref{sec:Cold-inst} and \ref{sec:Ultracold-inst}, in the cold and
ultracold case we have in addition an internal infinity boundary,
$\Sigma^{\rm int}_{\infty}$, for which the boundary action term
does not necessarily vanish.

The domain of validity of our results is the  particle limit,
$mA\ll 1$, for which the  radius of the black hole, $r_+ \sim m$,
is much smaller than the typical distance between the black holes
at the creation moment, $\ell \sim 1/A$.

In order to compute the black hole pair creation rate given by
(\ref{PC-rate-breakstring}) or (\ref{PC-rate}), we need to find
$I_{\rm string}$ and $I_{\rm dS}$. The evaluation of the action of
the string instanton that mediates the nucleation of a string in
the dS spacetime is done using
 (\ref{backg metric-dS}) and (\ref{delta phi0-dS}), yielding
\begin{eqnarray}
I_{\rm string}&=&-\frac{1}{16\pi}\int_{\cal{M}} d^4x\sqrt{g}
\left ( R-2\Lambda \right )\nonumber \\
&=& -\frac{3\pi}{2\Lambda}\frac{{\cal G}'(x_{\rm s})}{|{\cal
G}'(x_{\rm n})|} \,,
 \label{I-string}
 \end{eqnarray}
where the integration over $\tau$ has been done in the interval
$[0,\beta_{\rm 0}/2]$ with $\beta_{\rm
0}=2\pi/\sqrt{1+\Lambda/(3A_0^{\,2})}$, the integration range of
$x$ was $[-1,1]$, the integration interval of $y$ was
$[\sqrt{1+\Lambda/(3A_0^{\,2})}\,,\infty]$, and $R=4\Lambda$.

The action of the $S^4$ gravitational instanton that mediates the
nucleation of the dS spacetime is \cite{BoussoHawk,VolkovWipf}
\begin{eqnarray}
I_{\rm dS}=-\frac{3\pi}{2\Lambda} \,.
 \label{I-dS}
 \end{eqnarray}

The nucleation rate of a string in a dS background is then given
by (\ref{PC-rate-string}). For the particle limit, $mA\ll 1$, the
mass density of the string, $\mu$, is given by $\mu \simeq mA$ and
\begin{eqnarray}
\Gamma_{\rm string/dS} \sim e^{- 12\pi \frac{\mu}{\Lambda}} \:.
 \label{PC-rate-stringend}
 \end{eqnarray}
Thus, the nucleation probability of a string in the dS background
decreases when its mass density increases.
\subsection{\label{sec:Lukewarm-rate}The lukewarm C pair creation rate}

We first consider the magnetic case, whose Euclidean action is
given by (\ref{I}), and then we consider the electric case, using
(\ref{I-electric}), and we verify that these two quantities give
the same numerical value. The boundary $\Sigma=\partial \cal{M}$
that appears in (\ref{I}) consists of an initial spatial surface
at $\tau=0$ plus a final spatial surface at $\tau=\beta_{\rm
\ell}/2$. We label these two 3-surfaces by $\Sigma_{\tau}$. Each
one of these two spatial 3-surfaces is delimitated by a 2-surface
at the acceleration horizon and by a 2-surface at the outer black
hole horizon. The two surfaces $\Sigma_{\tau}$ are connected by a
timelike 3-surface that intersects $\Sigma_{\tau}$ at the frontier
$y_A$ and by a timelike 3-surface that intersects $\Sigma_{\tau}$
at the frontier $y_+$. We label these two timelike 3-surfaces by
$\Sigma_{h}$. Thus $\Sigma=\Sigma_{\tau}+\Sigma_{h}$, and the
region $\cal{M}$ within it is compact. With the analysis of
section \ref{sec:Lukewarm-inst}, we can compute all the terms of
action (\ref{I}). We start with
\begin{eqnarray}
& &\!\!\!\!\!\!\!\!\!\!\!\!\!\!-\frac{1}{16\pi}\int_{\cal{M}}
d^4x\sqrt{g} \left ( R-2\Lambda \right )=
       \nonumber \\
& & \!\!\!\!\!\!\!-\frac{1}{16\pi}\int_{\Delta \phi _{\rm \ell}}
\!\!\!\!d\phi
 \int_0^{\beta_{\rm \ell}/2} \!\!\!\!d\tau
 \int_{x_\mathrm{s}}^{x_\mathrm{n}}\!\!\!\! dx
 \int_{y_A}^{y_+} \!\!\!\!dy \frac{2 \Lambda}{\left [A(x+y) \right ]^4} \:,
 \label{I1-luk}
 \end{eqnarray}
where we have used $R=4\Lambda$, and $y_A$ and $y_+$ are given by
(\ref{yA-luk}), $x_\mathrm{s}$ and $x_\mathrm{n}$ are defined by
(\ref{polos-luk}), and $\beta_{\rm \ell}$ and $\Delta \phi _{\rm
\ell}$ are respectively given by  (\ref{beta-luk}) and
(\ref{Period phi-luk}). The Maxwell term in the action yields
 \begin{eqnarray}
 & & \!\!\!\!\!\!\!\!\! \frac{1}{16\pi}\int_{\cal{M}} \!\!d^4x\sqrt{g}
 \:F_{\rm mag}^2 \!=\! \frac{q^2}{16\pi}\,\Delta \phi _{\rm \ell}\, \beta_{\rm
\ell}\,(x_\mathrm{n}-x_\mathrm{s})(y_+ -y_A) \:, \nonumber \\
& &
 \label{I2-luk}
 \end{eqnarray}
where we have used $F_{\rm mag}^2=2q^2A^4(x+y)^4$ [see
(\ref{F-mag})], and $\int_{\Sigma} d^3x\sqrt{h}\, K=0$. Adding all
these terms yields for the magnetic action (\ref{I})
\begin{eqnarray}
\!\!\!\!\!\!\!\!\!\! I_{\rm mag}^{\rm
\ell}&=&-\frac{3\pi}{16\Lambda} \frac{1}{8mA}
 \left ( 1-\sqrt{\frac{1-4mA}{1+4mA}} \right ) \nonumber \\
 & & \!\!\!\!\! \times
  \left ( 1+\sqrt{1-(4mA)^2}-\frac{4m}{\sqrt{3}}\sqrt{\Lambda+3A^2}
   \right )\,,
\label{I-mag-luk}
 \end{eqnarray}
and, given that the string is already present in the initial
system, the pair creation rate of nonextreme lukewarm black holes
when the cosmic string breaks is
\begin{eqnarray}
\Gamma_{\rm BHs/string}^{\rm \ell}=\eta\,e^{-2I_{\rm mag}^{\rm
\ell}+2I_{\rm string}}\,, \label{PC rate-luk}
 \end{eqnarray}
where (\ref{I-string}) yields $I_{\rm string}=
-\frac{3\pi}{2\Lambda}\frac{\sqrt{1-4mA}}{\sqrt{1+4mA}}$, and
$\eta$ is the one-loop contribution not computed here.
 $I_{\rm mag}^{\rm \ell}$ is a monotonically increasing function of
both $m$ and $A$ (for a fixed $\Lambda$). When we take the limit
$A=0$ of (\ref{I-mag-luk}) we get
\begin{eqnarray}
I_{\rm mag}^{\rm \ell}{\biggl |}_{A\rightarrow
0}=-\frac{3\pi}{2\Lambda}+\pi \,m \sqrt{\frac{3}{\Lambda}} \,,
\label{I-mag-luk-A=0}
 \end{eqnarray}
recovering the action for the $A=0$ lukewarm instanton
\cite{MannRoss}, that describes the pair creation of two
non-extreme dS$-$Reissner-Nordstr\"{o}m black holes accelerated
only by the cosmological constant.

In the electric case, the Euclidean action is given by
(\ref{I-electric}) with  $F_{\rm el}^2=-2q^2A^4(x+y)^4$ [see
(\ref{F-el})]. Thus,
\begin{eqnarray}
\frac{1}{16\pi}\int_{\cal{M}} d^4x\sqrt{g}
 \:F_{\rm el}^2=- \frac{1}{16\pi}\int_{\cal{M}} d^4x\sqrt{g}
 \:F_{\rm mag}^2\:.
 \label{I2-luk-elect}
 \end{eqnarray}
In order to compute the extra Maxwell boundary term in
(\ref{I-electric}) we have to find a vector potential, $A_{\nu}$,
that is regular everywhere including at the horizons. An
appropriate choice in the lukewarm case is $A_y=- i\,q\,\tau$,
which obviously satisfies (\ref{F-el}). The integral over $\Sigma$
consists of an integration between $y_A$ and $y_+$ along the
$\tau=0$ surface and back along  $\tau=\beta_{\rm \ell}/2$, and of
an integration between $\tau=0$ and $\tau=\beta_{\rm \ell}/2$
along the $y=y_+$ surface and back along the $y=y_A$ surface. The
normal to $\Sigma_{\tau}$ is
$n_{\mu}=(\sqrt{{\cal{F}}}/[A(x+y)],0,0,0)$, and the normal to
$\Sigma_{h}$ is $n_{\mu}=(0,\sqrt{{\cal{F}}}/[A(x+y)],0,0)$. Thus
$F^{\mu\nu}n_{\mu}A_{\nu}=0$ on $\Sigma_{h}$, and the
non-vanishing contribution comes only from the integration along
the $\tau=\beta_{\rm \ell}/2$ surface. The Maxwell boundary term
in (\ref{I-electric}), $-\frac{1}{4\pi}\int_{\Sigma}
d^3x\sqrt{h}\, F^{\mu\nu}n_{\mu}A_{\nu}$, is then
 \begin{eqnarray}
& & \!\!\!\!\!\!\!\!\!\!\!\!\!\!\!\!\!\!\!\!\!\!\!\!\!\!\!\!\!
-\frac{1}{4\pi}\int_{\Sigma_{\tau=\beta_{\rm \ell}/2} } \!\!\!\!
d^3x \sqrt{g_{yy}g_{xx}g_{\phi\phi}}\:
F^{\tau y} n_{\tau} A_{y}=           \nonumber \\
& & \:\:\:\:\:\:\:\:\:\:
 \frac{q^2}{8\pi}\,\Delta \phi _{\rm
\ell}\, \beta_{\rm \ell}\,(x_\mathrm{n}-x_\mathrm{s})(y_+ -y_A)\:.
 \label{I-electric-luk}
 \end{eqnarray}
Adding  (\ref{I1-luk}), (\ref{I2-luk-elect}) and
(\ref{I-electric-luk}) yields for the electric action
(\ref{I-electric})
\begin{eqnarray}
 I_{\rm el}^{\rm \ell}= I_{\rm mag}^{\rm \ell}\,,
\label{I-elect-luk}
 \end{eqnarray}
where $I_{\rm mag}^{\rm \ell}$ is given by  (\ref{I-mag-luk}).

\begin{figure} [t]
\includegraphics*[height=2.1in]{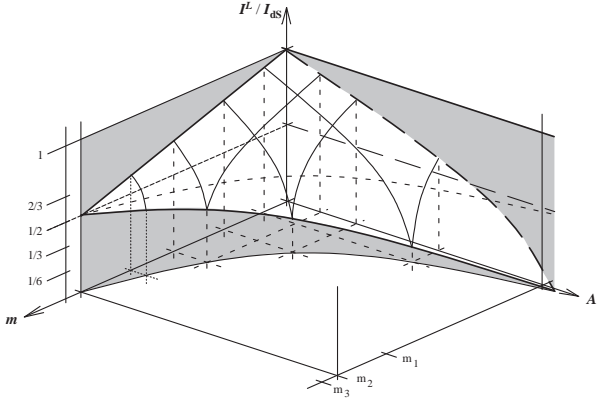}
\caption{\label{lukewarm-fig} Plot of $I^{\rm \ell}/I_{\rm dS}$ as
a function of $m$ and $A$ for a fixed $\Lambda$, where $I^{\rm
\ell}$ is the action of the lukewarm C instanton, given in
(\ref{I-mag-luk}), and $I_{\rm dS}=-\frac{3\pi}{2\Lambda}$ is the
action of de Sitter space.
 See text of Sec.  \ref{sec:Lukewarm-rate}. $m_1=1/(3\sqrt{\Lambda})$,
$m_2=\sqrt{3}/(4\sqrt{\Lambda})$ and
$m_3=\sqrt{2}/(3\sqrt{\Lambda})$.
 }
\end{figure}

 In Fig. \ref{lukewarm-fig} we show a
plot of $I^{\rm \ell}/I_{\rm dS}$ as a function of $m$ and $A$ for
a fixed $\Lambda$, where $I_{\rm dS}=-\frac{3\pi}{2\Lambda}$ is
the action of de Sitter space. Given the pair creation rate,
$\Gamma_{\rm BHs/string}^{\rm \ell}\propto e^{-2I_{\rm mag}^{\rm
\ell}+2I_{\rm string}}$, we conclude that, for a fixed $\Lambda$
and $A$, as the mass and charge of the lukewarm black holes
increase, the probability they have to be pair created decreases
monotonically. Moreover, for a fixed mass and charge, this
probability increases monotonically as the acceleration provided
by the string increases. Alternatively, we can discuss the
behavior of $\Gamma_{\rm BHs/dS}^{\rm \ell}\propto e^{-2I_{\rm
mag}^{\rm \ell}+2I_{\rm dS}}$. In this case, for a fixed mass and
charge, the probability decreases monotonically as the
acceleration of the black holes increases.

\subsection{\label{sec:Cold-rate}The cold C pair creation rate}
\begin{figure} [t]
\includegraphics*[height=2.1in]{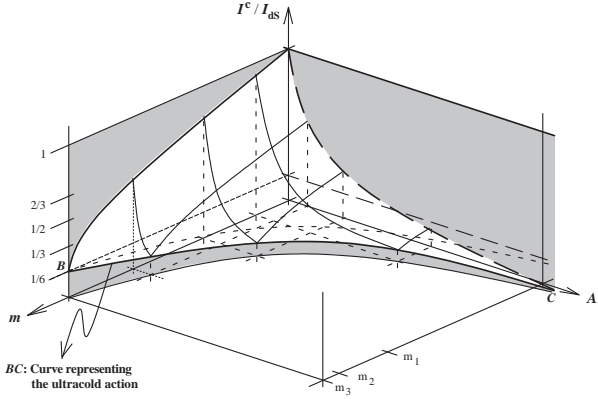}
\caption{\label{cold-fig}
 Plot of $I^{\rm c}/I_{\rm dS}$ as a
function of $m$ and $A$ for a fixed $\Lambda$, where $I^{\rm c}$
is the action of the cold C instanton, given in
(\ref{I-mag-cold}), and $I_{\rm dS}=-\frac{3\pi}{2\Lambda}$ is the
action of de Sitter space (see text of Sec. \ref{sec:Cold-rate}).
The plot for the ultracold C action, given in
(\ref{I-mag-ultracold}), is also represented by the curve $BC$
(see text of Sec.  \ref{sec:Ultracold-rate}).
$m_1=1/(3\sqrt{\Lambda})$, $m_2=\sqrt{3}/(4\sqrt{\Lambda})$ and
$m_3=\sqrt{2}/(3\sqrt{\Lambda})$.
 }
\end{figure}

We first consider the magnetic case, whose Euclidean action is
given by (\ref{I}). The boundary that appears in (\ref{I}) is
given by $\Sigma=\Sigma_{\tau}+\Sigma_{h}+\Sigma^{\rm
int}_{\infty}$, where $\Sigma_{\tau}$ is a spatial surface at
$\tau=0$ and $\tau=\beta_{\rm c}/2$, $\Sigma_{h}$ is a timelike
3-surface at $y=y_A$, and the timelike 3-surface $\Sigma^{\rm
int}_{\infty}$ is an internal infinity boundary at $y=y_+=\rho$.
With the analysis of Sec.  \ref{sec:Cold-inst}, we can compute all
the terms of action (\ref{I}). We start with
\begin{eqnarray}
& &\!\!\!\!\!\!\!\!\!\!\!\!\!\!-\frac{1}{16\pi}\int_{\cal{M}}
d^4x\sqrt{g} \left ( R-2\Lambda \right )=
       \nonumber \\
& & \!\!\!\!\!\!\!-\frac{1}{16\pi}\int_{\Delta \phi _{\rm c}}
\!\!\!\!d\phi
 \int_0^{\beta_{\rm c}/2} \!\!\!\!d\tau
 \int_{x_\mathrm{s}}^{x_\mathrm{n}}\!\!\!\! dx
 \int_{y_A}^{\rho} \!\!\!\!dy \frac{2 \Lambda}{\left [A(x+y) \right ]^4} \:,
 \label{I1-cold}
 \end{eqnarray}
where we have used $R=4\Lambda$, and $\rho$ and $y_A$ are
respectively given by (\ref{zerosy1-cold}) and
(\ref{zerosy3-cold}), $x_\mathrm{s}$ and $x_\mathrm{n}$ are
defined by (\ref{polos-cold}), and $\beta_{\rm c}$ and $\Delta
\phi _{\rm c}$ are, respectively, given by (\ref{beta-cold}) and
(\ref{Period phi}). The Maxwell term in the action yields
 \begin{eqnarray}
& & \!\!\!\!\!\!\!\!\!\!\! \frac{1}{16\pi}\int_{\cal{M}} \!\!
d^4x\sqrt{g}
 \:F_{\rm mag}^2 \!=\! \frac{q^2}{16\pi}\,\Delta \phi _{\rm c}\, \beta_{\rm
c}\,(x_\mathrm{n}-x_\mathrm{s})(\rho -y_A) \:, \nonumber \\
& &
 \label{I2-cold}
 \end{eqnarray}
where we have used $F_{\rm mag}^2=2q^2A^4(x+y)^4$ [see
(\ref{F-mag})], and $\int_{\Sigma} d^3x\sqrt{h}\, K=0$. Adding all
these terms yields for the magnetic action (\ref{I}) of the cold
case
\begin{eqnarray}
I_{\rm mag}^{\rm c}=-\frac{\Delta \phi _{\rm c}} {8A^2}\,
 \frac{ x_\mathrm{n}-x_\mathrm{s} }
   {(x_\mathrm{n}+y_A)(x_\mathrm{s}+y_A)}\,.
    \label{I-mag-cold}
 \end{eqnarray}
Given that the string is already present in the initial system,
the pair creation rate of extreme cold black holes when the string
breaks is $\Gamma_{\rm BHs/string}^{\rm c}=\eta\,e^{-2I_{\rm
mag}^{\rm c}+2I_{\rm string}}$, where $I_{\rm string}$ is given by
(\ref{I-string}), and $\eta$ is the one-loop contribution not
computed here. In Fig. \ref{cold-fig} we show a plot of $I_{\rm
mag}^{\rm c}/I_{\rm dS}$ as a function of $m$ and $A$ for a fixed
$\Lambda$. Given the pair creation rate, $\Gamma_{\rm
BHs/string}^{\rm c}$, we conclude that for a fixed $\Lambda$ and
$A$ as the mass and charge of the cold black holes increases, the
probability they have to be pair created decreases monotonically.
Moreover, for a fixed mass and charge, this probability increases
monotonically as the acceleration of the black holes increases.
Alternatively, we can discuss the behavior of $\Gamma_{\rm
BHs/dS}^{\rm c}\propto e^{-2I_{\rm mag}^{\rm c}+2I_{\rm dS}}$. In
this case, for a fixed mass and charge, the probability decreases
monotonically as the acceleration of the black holes increases.
 When we take the limit $A=0$ we recover the action for the $A=0$
cold instanton \cite{MannRoss}, which lies in the range
$-\frac{3\pi}{2\Lambda} \leq I_{\rm mag}^{\rm c}{\bigl
|}_{A\rightarrow 0}\leq -\frac{\pi}{4\Lambda}$, and which
describes the pair creation of extreme dS$-$Reissner-Nordstr\"{o}m
black holes accelerated only by the cosmological constant.

In the electric case, the Euclidean action is given by
(\ref{I-electric}) with  $F_{\rm el}^2=-2q^2A^4(x+y)^4$ [see
(\ref{F-el})]. Thus,
\begin{eqnarray}
\frac{1}{16\pi}\int_{\cal{M}} d^4x\sqrt{g}
 \:F_{\rm el}^2=- \frac{1}{16\pi}\int_{\cal{M}} d^4x\sqrt{g}
 \:F_{\rm mag}^2\:.
 \label{I2-cold-elect}
 \end{eqnarray}
In order to compute the extra Maxwell boundary term in
(\ref{I-electric}) we have to find a vector potential, $A_{\nu}$,
that is regular everywhere including at the horizons. An
appropriate choice in the cold case is $A_y=- i\,q\,\tau$, which
obviously satisfies (\ref{F-el}). Analogously to the lukewarm
case, the non-vanishing contribution to the Maxwell boundary term
in (\ref{I-electric}) comes only from the integration along the
$\tau=\beta_{\rm c}/2$ surface, and is given by
 \begin{eqnarray}
& & \!\!\!\!\!\!\!\!\!\!\!\!\!\!\!\!\!\!\!\!\!\!\!\!\!\!\!\!\!
-\frac{1}{4\pi}\int_{\Sigma_{\tau=\beta_{\rm c}/2}} \!\!\!\! d^3x
\sqrt{g_{yy}g_{xx}g_{\phi\phi}}\:
F^{\tau y} n_{\tau} A_{y}=           \nonumber \\
& & \:\:\:\:\:\:\:\:\:\: \frac{q^2}{8\pi}\,\Delta \phi _{\rm
c}\,\beta_{\rm c}\,(x_\mathrm{n}-x_\mathrm{s})(\rho -y_A)\:.
 \label{I-electric-cold}
 \end{eqnarray}
Adding (\ref{I2-cold-elect}) and (\ref{I-electric-cold}) yields
(\ref{I2-cold}). Thus, the electric action (\ref{I-electric}) of
the cold instanton is equal to the magnetic action, $I_{\rm
el}^{\rm c}= I_{\rm mag}^{\rm c}$, and therefore electric and
magnetic cold black holes have the same probability of being pair
created.

\subsection{\label{sec:Nariai-rate}The Nariai C pair creation rate}
The Nariai C instanton is the only one that can have zero charge.
We will first consider the charged Nariai C instanton and then the
neutral Nariai C instanton.

\begin{figure}[t]
\includegraphics*[height=2.1in]{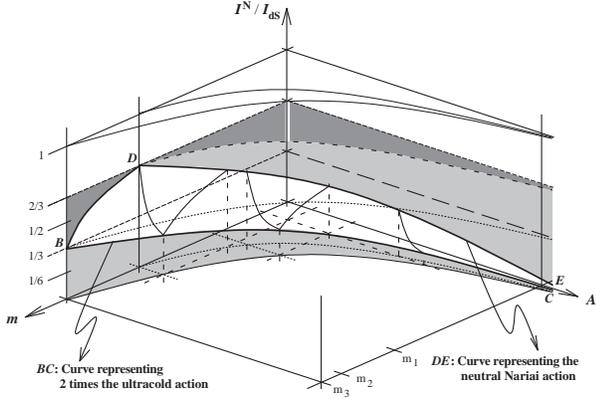}
\caption{\label{nariai-fig-PCAdS}
 Plot of $I^{\rm N}/I_{\rm dS}$ as a
function of $m$ and $A$ for a fixed $\Lambda$, where $I^{\rm N}$
is the action of the cold C instanton, given in
(\ref{I-mag-Nariai}), and $I_{\rm dS}=-\frac{3\pi}{2\Lambda}$ is
the action of de Sitter space. The neutral Nariai plot is sketched
by the curve $DE$ (see text of Sec.  \ref{sec:Nariai-rate}). The
curve $BC$ represents two times the ultracold C action, given in
(\ref{I-mag-ultracold}). $m_1=1/(3\sqrt{\Lambda})$,
$m_2=\sqrt{3}/(4\sqrt{\Lambda})$ and
$m_3=\sqrt{2}/(3\sqrt{\Lambda})$.
 }
\end{figure}

We start with the magnetic case, whose Euclidean action is given
by (\ref{I}). The boundary that appears in (\ref{I}) is given by
$\Sigma=\Sigma_{\tilde{\tau}}+\Sigma_{h}$, where
$\Sigma_{\tilde{\tau}}$ is a spatial surface at $\tilde{\tau}=0$
and $\tilde{\tau}=\pi$, and $\Sigma_{h}$ is a timelike 3-surface
at $\chi=0$ and $\chi=\pi$. With the analysis of Sec.
\ref{sec:Nariai-inst}, we can compute all the terms of action
(\ref{I}). We start with
\begin{eqnarray}
& &\!\!\!\!\!\!\!\!\!\!\!\!\!\!-\frac{1}{16\pi}\int_{\cal{M}}
d^4x\sqrt{g} \left ( R-2\Lambda \right )=
       \nonumber \\
& & \!\!\!\!\!\!\!\!\!\!
  -\frac{1}{16\pi}\int_{\Delta \phi _{\rm N}} \!\!\!\!d\phi
 \int_0^{\pi} \!\!\!\!d\tilde{\tau}
 \int_{x_\mathrm{s}}^{x_\mathrm{n}}\!\!\!\! dx
 \int_{0}^{\pi} \!\!\!\!d\chi \:
 \frac{2 \Lambda \sin \chi}{\left [A(x+\rho) \right ]^4 {\cal{K}} }\:,
 \label{I1-nariai}
 \end{eqnarray}
where we have used $R=4\Lambda$, $x_\mathrm{s}$ and $x_\mathrm{n}$
are defined by (\ref{polos-cold}), and $\Delta \phi _{\rm N}$ is
given by (\ref{Period phi}). The Maxwell term in the action yields
\begin{eqnarray}
 \frac{1}{16\pi}\int_{\cal{M}} d^4x\sqrt{g}
 \:F_{\rm mag}^2 =\frac{q^2}{4 \, \cal{K}}\,\Delta \phi _{\rm N}\,(x_\mathrm{n}-x_\mathrm{s}) \:,
 \label{I2-nariai}
 \end{eqnarray}
where we have used $F_{\rm mag}^2=2q^2A^4(x+\rho)^4$ [see
(\ref{F-mag})], and $\int_{\Sigma} d^3x\sqrt{h}\, K=0$. Adding
these three terms yields the magnetic action (\ref{I}) of the
Nariai case
\begin{eqnarray}
I_{\rm mag}^{\rm N}=
 -\,\frac{\Delta \phi _{\rm N}} {4A^2}\,
 \frac{ x_\mathrm{n}-x_\mathrm{s} }
   {(x_\mathrm{n}+\rho)(x_\mathrm{s}+\rho)}\,,
    \label{I-mag-Nariai}
 \end{eqnarray}
 where  $m$ and $q$ are subjected to
(\ref{mq-cNariai PCdS}). Given that the string is already present
in the initial system, the pair creation rate of extreme Nariai
black holes when the string breaks is $\Gamma_{\rm
BHs/string}^{\rm N}=\eta\,e^{-2I_{\rm mag}^{\rm N}+2I_{\rm
string}}$, where $I_{\rm string}$ is given by (\ref{I-string}),
and $\eta$ is the one-loop contribution not computed here. In Fig.
\ref{nariai-fig-PCAdS} we show a plot of $I_{\rm mag}^{\rm
N}/I_{\rm dS}$ as a function of $m$ and $A$ for a fixed $\Lambda$.
Given the pair creation rate, $\Gamma_{\rm BHs/string}^{\rm N}$,
we conclude that for a fixed $\Lambda$ and $A$ as the mass and
charge of the Nariai black holes increases, the probability they
have to be pair created decreases monotonically. Moreover, for a
fixed mass and charge, this probability increases monotonically as
the acceleration of the black holes increases. Alternatively, we
can discuss the behavior of $\Gamma_{\rm BHs/dS}^{\rm N}\propto
e^{-2I_{\rm mag}^{\rm N}+2I_{\rm dS}}$. In this case, for a fixed
mass and charge, the probability decreases monotonically as the
acceleration of the black holes increases.
 When we take the limit $A=0$ we recover the action for the $A=0$
Nariai instanton \cite{MannRoss,HawkRoss}, which lies in the range
$-\frac{\pi}{\Lambda} \leq I_{\rm mag}^{\rm N}{\bigl
|}_{A\rightarrow 0}\leq -\frac{\pi}{2\Lambda}$, and that describes
the nucleation of a Nariai universe that is unstable
\cite{GinsPerry,BoussoHawk,Bousso60y} and decays through the pair
creation of extreme dS$-$Reissner-Nordstr\"{o}m black holes
accelerated only by the cosmological constant.

In the electric case, the Euclidean action is given by
(\ref{I-electric}) with  $F_{\rm el}^2=-2q^2A^4(x+\rho)^4$ [see
(\ref{F-el-Nariai PCdS})]. Thus,
\begin{eqnarray}
\frac{1}{16\pi}\int_{\cal{M}} d^4x\sqrt{g}
 \:F_{\rm el}^2=- \frac{1}{16\pi}\int_{\cal{M}} d^4x\sqrt{g}
 \:F_{\rm mag}^2\:.
 \label{I2-nariai-elect}
 \end{eqnarray}
In order to compute the extra Maxwell boundary term in
(\ref{I-electric}), the appropriate  vector potential, $A_{\nu}$,
that is regular everywhere including at the horizons is
$A_{\chi}=i\,\frac{q}{\cal{K}}\,\tilde{\tau}\, \sin \chi$, which
obviously satisfies (\ref{F-el-Nariai PCdS}). The integral over
$\Sigma$ consists of an integration between $\chi=0$ and
$\chi=\pi$ along the $\tilde{\tau}=0$ surface and back along
$\tilde{\tau}=\pi$, and of an integration between $\tilde{\tau}=0$
and $\tilde{\tau}=\pi$ along the $\chi=0$ surface, and back along
the $\chi=\pi$ surface.
 The unit normal to $\Sigma_{\tilde{\tau}}$ is
$n_{\mu}=(\frac{\sin \chi}{\sqrt{\cal{K}}A(x+\rho)},0,0,0)$, and
$F^{\mu\nu}n_{\mu}A_{\nu}=0$ on $\Sigma_{h}$. Therefore, the
non-vanishing contribution to the Maxwell boundary term in
(\ref{I-electric}), $-\frac{1}{4\pi}\int_{\Sigma} d^3x\sqrt{h}\,
F^{\mu\nu}n_{\mu}A_{\nu}$, comes only from the integration along
the $\tilde{\tau}=\pi$ surface and is given by
 \begin{eqnarray}
-\frac{1}{4\pi}\int_{\Sigma_{\tilde{\tau}=\pi} } \!\!\!\!\!\! d^3x
\sqrt{h}\: F^{\tilde{\tau} \chi} n_{\tilde{\tau}} A_{\chi}=
\frac{q^2}{2 \, \cal{K}}\,\Delta \phi _{\rm
N}\,(x_\mathrm{n}-x_\mathrm{s})\:.
 \label{I-electric-nariai}
 \end{eqnarray}
Adding (\ref{I2-nariai-elect}) and (\ref{I-electric-nariai})
yields (\ref{I2-nariai}). So, the electric action
(\ref{I-electric}) of the charged Nariai instanton is equal to the
magnetic action, $I_{\rm el}^{\rm N}= I_{\rm mag}^{\rm N}$, and
therefore electric and magnetic charged Nariai black holes have
the same probability of being pair created.

Now, we discuss the neutral Nariai C instanton. This instanton is
particulary important since it is the only regular Euclidean
solution available when we want to evaluate the pair creation of
neutral black holes. The same feature is present in the $A=0$ case
where only the neutral Nariai instanton is available
\cite{GinsPerry,MannRoss,BoussoHawk,VolkovWipf}. The action of the
neutral Nariai C instanton is simply given by (\ref{I1-nariai})
and, for a fixed $\Lambda$ and $A$, it is always smaller than the
action of the charged Nariai C instanton (see line $DE$ in Fig.
\ref{nariai-fig-PCAdS}): $I_{\rm charged}^{\rm N}> I_{\rm
neutral}^{\rm N}>I_{\rm dS}$. Thus the pair creation of charged
Nariai black holes is suppressed relative to the pair creation of
neutral Nariai black holes, and both are suppressed relative to
the dS space.

\subsection{\label{sec:Ultracold-rate}The ultracold C pair creation rate}
We first consider the magnetic case, whose Euclidean action is
given by (\ref{I}). The boundary that appears in (\ref{I}) is
given by $\Sigma=\Sigma_{\tilde{\tau}}+\Sigma_{h}+\Sigma^{\rm
int}_{\infty}$, where $\Sigma_{\tilde{\tau}}$ is a spatial surface
at $\tilde{\tau}=0$ and $\tilde{\tau}=\pi$, $\Sigma_{h}$ is a
timelike 3-surface at the Rindler horizon $\chi=0$, and the
timelike 3-surface $\Sigma^{\rm int}_{\infty}$ is an internal
infinity boundary at $\chi=\infty$. With the analysis of Sec.
\ref{sec:Ultracold-inst}, we can compute all the terms of action
(\ref{I}). We start with $-\frac{1}{16\pi}\int_{\cal{M}}
d^4x\sqrt{g} \left ( R-2\Lambda \right )$ which yields  (using
$R=4\Lambda$)
\begin{eqnarray}
& &\!\!\!\!\!\!\!\!\!\!\!\!\!\!\!\!\!\!\!\!\!\!\!\!\!
  -\frac{1}{16\pi}\int_{\Delta \phi _{\rm u}} \!\!\!\!d\phi
 \int_0^{\pi} \!\!\!\!d\tilde{\tau}
 \int_{x_\mathrm{s}}^{x_\mathrm{n}}\!\!\!\! dx
 \int_{0}^{\chi_0} \!\!\!\!d\chi \:
 \frac{2 \Lambda \chi}{(Ax+\rho)^4 }=
       \nonumber \\
& &
  -\frac{\Lambda}{16}\:\Delta \phi _{\rm u} \:\chi_0^2 \:
  \int_{x_\mathrm{s}}^{x_\mathrm{n}}\!\!\!\! dx
 \frac{1}{(Ax+A\rho)^4 }
  {\biggl |}_{\chi_0\rightarrow \infty}\:,
 \label{I1-ultracold}
 \end{eqnarray}
where  $x_\mathrm{s}$ and $x_\mathrm{n}$ are defined by
(\ref{polos-cold}) and (\ref{mq-ultra}), and $\Delta \phi _{\rm
u}$ is given by (\ref{Period phi}). The Maxwell term in the action
yields
\begin{eqnarray}
 \frac{1}{16\pi}\int_{\cal{M}} d^4x\sqrt{g}
 \:F_{\rm mag}^2 =\frac{q^2}{16}\,\Delta \phi _{\rm
u}\:\chi_0^2\,(x_\mathrm{n}-x_\mathrm{s})
  {\biggl |}_{\chi_0\rightarrow \infty}\!\!\!\!,
 \label{I2-ultracold}
 \end{eqnarray}
where we have used $F_{\rm mag}^2=2q^2A^4(x+\rho)^4$ [see
(\ref{F-mag})] with $\rho=\sqrt{2(\Lambda+3A^2)}$. Due to the fact
that $\chi_0\rightarrow \infty$ it might seem that the
contribution from (\ref{I1-ultracold}) and (\ref{I2-ultracold})
diverges. Fortunately this is not the case since these two terms
cancel each other. Trying to verify this analytically is
cumbersome, but for our purposes we can simply fix any numerical
value for $\Lambda$ and $A$, and using (\ref{mq-ultra}) and
(\ref{polos-cold}) we indeed verify that (\ref{I1-ultracold}) and
(\ref{I2-ultracold}) cancel each other.

Now, contrary to the other instantons, the ultracold C instanton
has a non-vanishing extrinsic curvature boundary term,
$-\frac{1}{16\pi}\int_{\Sigma} d^3x\sqrt{h}\, K \neq 0$, due to
the internal infinity boundary ($\Sigma^{\rm int}_{\infty}$ at
$\chi=\infty$) contribution. The extrinsic curvature to
$\Sigma^{\rm int}_{\infty}$ is
$K_{\mu\nu}=h_{\mu}^{\:\:\:\alpha}\nabla_{\alpha}n_{\nu}$, where
 $n_{\nu}=(0,\frac{1}{A(x+\rho)},0,0)$ is the unit outward normal to
$\Sigma^{\rm int}_{\infty}$,
$h_{\mu}^{\:\:\:\alpha}=g_{\mu}^{\:\:\:\alpha}-n_{\mu}n^{\alpha}
 =(1,0,1,1)$ is the projection tensor onto $\Sigma^{\rm int}_{\infty}$,
 and $\nabla_{\alpha}$ represents the covariant derivative with respect
 to $g_{\mu\nu}$. Thus the trace of the extrinsic
curvature to $\Sigma^{\rm int}_{\infty}$ is
$K=g^{\mu\nu}K_{\mu\nu}=A(x+\rho)/\chi$, and
\begin{eqnarray}
 & & \!\!\!\!\!\!\!\!\!\!\!\!\!\!
  -\frac{1}{8\pi}\int_{\Sigma} d^3x\sqrt{h}\, K=
 \nonumber \\
 & &-\frac{1}{8\pi}\,\int_{\Delta \phi _{\rm u}} \!\!\!\!d\phi
 \int_0^{\pi} \!\!\!\!d\tilde{\tau}
 \int_{x_\mathrm{s}}^{x_\mathrm{n}}\!\!\!\! dx \:
 \frac{1}{(Ax+A\rho)^2}.
 \label{I3-ultracold}
 \end{eqnarray}
The magnetic action (\ref{I}) of the ultracold C instanton is then
\begin{eqnarray}
 & & \!\!\!\!\!\!\!\! I_{\rm mag}^{\rm u}=
 \nonumber \\
 & & \!\!\!\!\!-\frac{\pi}{4}
 \left [ x_\mathrm{n} \left ( 1+A\sqrt{\frac{2}{\Lambda+3A^2}}\,x_\mathrm{n}
 +\frac{A^2}{2(\Lambda+3A^2)}x_\mathrm{n}^2 \right ) \right ]^{-1}
  \nonumber \\
 & & \!\!\!\!\! \times
  \frac{x_\mathrm{n}-x_\mathrm{s}}
  {\left[ A x_\mathrm{n}+\sqrt{2(\Lambda+3A^2)}\right ]
   \left[ A x_\mathrm{s}+\sqrt{2(\Lambda+3A^2)}\right ]}\,,
\label{I-mag-ultracold}
 \end{eqnarray}
 where  $x_\mathrm{s}$ and $x_\mathrm{n}$ are defined by
(\ref{polos-cold}) and (\ref{mq-ultra}). When we take the limit
$A=0$ we get $x_\mathrm{s}=-1$ and $x_\mathrm{n}=1$, and
\begin{eqnarray}
I_{\rm mag}^{\rm u}{\biggl |}_{A\rightarrow 0}
=-\frac{\pi}{4\Lambda}\,,
 \label{I-mag-ultracold-A=0}
 \end{eqnarray}
and therefore we recover the action for the $A=0$ ultracold
instanton \cite{MannRoss}, that describes the pair creation of
ultracold black holes accelerated only by the cosmological
constant.

 In Fig. \ref{lukewarm-fig} we show a plot of $I_{\rm mag}^{\rm
u}/I_{\rm dS}$ as a function of $m$ and $A$ for a fixed $\Lambda$.
When we fix $\Lambda$ and $A$ we also fix the mass and charge of
the ultracold black holes. For a fixed $\Lambda$, when $A$
increases the probability of pair creation of ultracold black
holes, $\Gamma_{\rm BHs/string}^{\rm u}$, increases monotonically
and they have a lower mass and charge. Alternatively, we can
discuss the behavior of $\Gamma_{\rm BHs/dS}^{\rm u}$. In this
case, the probability decreases monotonically as the acceleration
of the black holes increases.

In the electric case, the Euclidean action is given by
(\ref{I-electric}) with  $F_{\rm el}^2=-2q^2A^4(x+\rho)^4$ [see
 (\ref{F-el-ultracold})]. Thus,
\begin{eqnarray}
\frac{1}{16\pi}\int_{\cal{M}} d^4x\sqrt{g}
 \:F_{\rm el}^2=- \frac{1}{16\pi}\int_{\cal{M}} d^4x\sqrt{g}
 \:F_{\rm mag}^2\:.
 \label{I2-ultracold-elect}
 \end{eqnarray}
In the ultracold case the vector potential $A_{\nu}$, that is
regular everywhere including at the horizon, needed to compute the
extra Maxwell boundary term in (\ref{I-electric}) is
$A_{\tilde{\tau}}=i\,\frac{q}{2}\, \chi^2$, which obviously
satisfies (\ref{F-el-ultracold}). The integral over $\Sigma$
consists of an integration between $\chi=0$ and $\chi=\infty$
along the $\tilde{\tau}=0$ surface and back along
$\tilde{\tau}=\pi$, and of an integration between $\tilde{\tau}=0$
and $\tilde{\tau}=\pi$ along the $\chi=0$ surface, and back along
the internal infinity
 surface $\chi=\infty$. The non-vanishing contribution to the Maxwell
boundary term in (\ref{I-electric}) comes only from the
integration along the internal infinity boundary $\Sigma^{\rm
int}_{\infty}$, and is given by
 \begin{eqnarray}
& & \!\!\!\!\!\!\!\!\!\!\!\!\!\!\!\!\!\!\!\!\!\!\!\!\!\!\!\!\!
-\frac{1}{4\pi}\int_{\Sigma^{\rm int}_{\infty}}  d^3x
\sqrt{g_{\tilde{\tau}\tilde{\tau}}g_{xx}g_{\phi\phi}}\:
F^{\chi\tilde{\tau}} n_{\chi} A_{\tilde{\tau}}=           \nonumber \\
& & \:\:\:\:\:\:\:\:\:\:\:\:\:\:\:\:\:\:\:\: \frac{q^2}{8}\,\Delta
\phi _{\rm u}\:\chi_0^2\,(x_\mathrm{n}-x_\mathrm{s})
  {\biggl |}_{\chi_0\rightarrow \infty}\:.
 \label{I-electric-ultracold}
 \end{eqnarray}
Adding (\ref{I2-ultracold-elect}) and (\ref{I-electric-ultracold})
yields (\ref{I2-ultracold}). Thus, the electric action
(\ref{I-electric}) of the ultracold C instanton is equal to the
magnetic action, $I_{\rm el}^{\rm u}= I_{\rm mag}^{\rm u}$, and
therefore electric and magnetic ultracold black holes have the
same probability of being pair created.

\subsection{\label{sec:Sub-Maximal-rate}Pair creation rate of nonextreme sub-maximal black holes}

The lukewarm, cold, Nariai and ultracold C-metric instantons are
saddle point solutions free of conical singularities both in the
$y_+$ and $y_A$ horizons. The corresponding black holes may then
nucleate in the dS background when a cosmic string breaks, and we
have computed their pair creation rates in the last four
subsections. However, these particular black holes are not the
only ones that can be pair created. Indeed, it has been shown in
\cite{WuSubMax,BoussoHawkSubMax} that Euclidean solutions with
conical singularities may also be used as saddle points for the
pair creation process. In this way, pair creation of nonextreme
sub-maximal black holes is allowed (by this nomenclature we mean
all the nonextreme black holes other than the lukewarm ones that
are in the region interior to the close line $NOUN$ in Fig.
\ref{mq-fig}), and their pair creation rate may be computed. In
order to calculate this rate, the action is given by (\ref{I}) and
(\ref{I-electric}) (in the magnetic and electric cases
respectively) and, in addition, it has now an extra contribution
from the conical singularity (c.s.) that is present in one of the
horizons ($y_+$, say) given by
\cite{ReggeGibbonsPerryAconSing,GinsPerry}
\begin{eqnarray}
\frac{1}{16\pi}\int_{\cal{M}} d^4x\sqrt{g}
 \:\left ( R-2\Lambda \right ){\biggl |}_{{\rm c.s.}\:{\rm at}\:y_+}
 \!\!\!= \frac{ {\cal A}_+\:\delta}{16\pi}\:,
 \label{I conical sing}
 \end{eqnarray}
where ${\cal A}_+=\int_{y=y_+} \sqrt{g_{xx}g_{\phi\phi}}\: dx
\,d\phi$ is the area of the 2-surface spanned by the conical
singularity, and
\begin{eqnarray}
\delta=2\pi \left ( 1-\frac{\beta_A}{\beta_+}\right )
 \label{delta concical sing PCdS}
 \end{eqnarray}
is the deficit angle associated to the conical singularity at the
horizon $y_+$, with $\beta_A=4 \pi / |{\cal F}'(y_A)|$ and
$\beta_+=4 \pi / |{\cal F}'(y_+)|$ being the periods of $\tau$
that avoid a conical singularity in the horizons $y_A$ and $y_+$,
respectively. The contribution from (\ref{I}) and
(\ref{I-electric}) follows straightforwardly in a similar way as
the one shown in subsection \ref{sec:Lukewarm-rate} with the
period of $\tau$, $\beta_A$,  chosen in order to avoid the conical
singularity at the acceleration horizon, $y=y_A$. The full
Euclidean action for general nonextreme sub-maximal black holes is
then
\begin{eqnarray}
& & \!\!\!\!\!\!\!\! I\!=\!
 \frac{\Delta \phi} {A^2}\,\left ( \!
 \frac{ x_\mathrm{n}-x_\mathrm{s} }
   {(x_\mathrm{n}+y_A)(x_\mathrm{s}+y_A)}
  \! + \!\frac{ x_\mathrm{n}-x_\mathrm{s} }
   {(x_\mathrm{n}+y_+)(x_\mathrm{s}+y_+)}\!
   \right ), \nonumber \\
& & \label{I-Sub-Maximal}
 \end{eqnarray}
where $\Delta \phi$ is given by (\ref{Period phi}), and the pair
creation rate of nonextreme sub-maximal  black holes is given by
(\ref{PC-rate-breakstring}) or (\ref{PC-rate}) with the use of
(\ref{I-string}) and (\ref{I-dS}). In order to compute
(\ref{I-Sub-Maximal}), we need the relation between the parameters
$A$, $\Lambda$, $m$, $q$, and the horizons $y_A$, $y_+$ and $y_-$.
In general, for a nonextreme solution with horizons $y_A<y_+<y_-$,
one has
\begin{eqnarray}
{\cal F}(y)= -\frac{1}{\mu}(y-y_A)(y-y_+)(y-y_-)(ay+b) \:,
 \label{F-nonext-sub-max}
 \end{eqnarray}
with
\begin{eqnarray}
\mu\!\!&=&\!\! y_A y_+ y_- (y_A +y_+ +y_-) +(y_A y_+ +y_A y_- +y_+
y_-)^2 \nonumber \\
 a\!\!&=&\!\!
 \left ( y_A y_+ +y_A y_- +y_+y_- \right )  \nonumber \\
 b\!\!&=&\!\! y_A y_+y_-\:.
 \label{F-nonext-sub-max aux}
 \end{eqnarray}
The parameters $A$, $\Lambda$, $m$ and $q$ can be expressed as a
function of $y_A$, $y_+$ and $y_-$ by
\begin{eqnarray}
\frac{\Lambda}{3A^2}\!\!&=&\!\! \mu^{-2}(y_A y_+y_-)^2 -1 \nonumber \\
 q^2A^2\!\!&=&\!\! \mu^{-1}(y_A y_+ +y_A y_-
+y_+ y_-) \nonumber \\
 mA\!\!&=&\!\! (2\,\sigma)^{-1}(y_A +y_+)(y_A +y_-)(y_+ +y_-) \nonumber \\
 \sigma\!\!&=&\!\! y_A^2 y_+y_- +y_A y_+^2 y_- +y_A y_+ y_-^2+ (y_A y_+)^2 \nonumber \\
  & & +(y_A y_-)^2 +(y_+ y_-)^2 \:.
 \label{relation parameters nonext-sub-max}
 \end{eqnarray}
The allowed values of parameters $m$ and $q$ are those contained
in the interior region defined by the close line $NOUN$ in Fig.
\ref{mq-fig}.

\section{\label{sec:Entropy}Entropy, area and pair creation rate}

In previous works on black hole pair creation in general
background fields it has been well established that the pair
creation rate is proportional to the exponential of the
gravitational entropy $S$ of the system, $\Gamma \propto e^S$,
with the entropy being given by one quarter of the the total area
$\cal{A}$ of all the horizons present in the instanton,
$S={\cal{A}}/4$. In what follows we will verify that these
relations also hold for the instantons of the dS C-metric.

\subsection{\label{sec:Lukewarm-S}The lukewarm C case. Entropy and area}
In the lukewarm case, the instanton has two horizons in its
Euclidean section, namely the acceleration horizon at $y=y_A$ and
the black hole horizon at $y=y_+$. So, the total area of the
lukewarm C instanton is
\begin{eqnarray}
\cal{A}^{\rm \ell}\!\! &=&\!\! \int_{y=y_A} \!\!\!
\sqrt{g_{xx}g_{\phi\phi}}\: dx \,d\phi + \int_{y=y_+} \!\!\!
\sqrt{g_{xx}g_{\phi\phi}}\:
dx \,d\phi =          \nonumber \\
& &\!\!\!\!\!\!\!\!\!\!\!\!
 \frac{\Delta \phi _{\rm \ell}} {A^2}\,\left (
 \frac{ x_\mathrm{n}-x_\mathrm{s} }
   {(x_\mathrm{n}+y_A)(x_\mathrm{s}+y_A)}
   +\frac{ x_\mathrm{n}-x_\mathrm{s} }
   {(x_\mathrm{n}+y_+)(x_\mathrm{s}+y_+)}
   \right )\,, \nonumber \\
& &
 \label{area-luk}
 \end{eqnarray}
 where $y_A$ and $y_+$ are given by (\ref{yA-luk}), $x_\mathrm{s}$
and $x_\mathrm{n}$ are defined by (\ref{polos-luk}), and
 $\Delta \phi _{\rm \ell}$ is
given by (\ref{Period phi-luk}). It is straightforward to verify
that ${\cal{A}^{\rm \ell}}=-8I^{\rm \ell}$, where $I^{\rm \ell}$
is given by (\ref{I-mag-luk}), and thus $\Gamma^{\rm \ell} \propto
e^{S^{\rm \ell}}$, where $S^{\rm \ell}= {\cal{A}^{\rm \ell}}/4$.

\subsection{\label{sec:Cold-S}The cold C case. Entropy and area}

In the cold case, the instanton has a single horizon,  the
acceleration horizon at $y=y_A$, in its Euclidean section, since
 $y=y_+$ is an internal infinity. So, the total area of the
cold C instanton is
\begin{eqnarray}
& &\!\!\!\!\!{\cal A}^{\rm c}\!=
 \int_{y=y_A}\!\!\! \sqrt{g_{xx}g_{\phi\phi}}\: dx \,d\phi \!=\!
 \frac{\Delta \phi _{\rm c}} {A^2}\,
 \frac{ x_\mathrm{n}-x_\mathrm{s} }
   {(x_\mathrm{n}+y_A)(x_\mathrm{s}+y_A)}\,, \nonumber \\
& &
 \label{area-cold}
 \end{eqnarray}
 where $y_A$ is given by (\ref{zerosy3-cold}), $x_\mathrm{s}$
and $x_\mathrm{n}$ are defined by (\ref{polos-cold}), and
 $\Delta \phi _{\rm c}$ is
given by (\ref{Period phi}). Thus, ${\cal{A}^{\rm c}}=-8I^{\rm
c}$, where $I^{\rm c}$ is given by (\ref{I-mag-cold}), and thus
$\Gamma^{\rm c} \propto e^{S^{\rm c}}$, where $S^{\rm c}=
{\cal{A}^{\rm c}}/4$.

\subsection{\label{sec:Nariai-S}The Nariai C case. Entropy and area}

In the Nariai case, the instanton has two horizons in its
Euclidean section, namely the acceleration horizon $y_A$ and the
black hole horizon $y_+$, both at $y=\rho$, and thus they have the
same area. So, the total area of the Nariai C instanton is
\begin{eqnarray}
& &\!\!\!{\cal A}^{\rm N}\!\!=\! 2\int_{y=\rho}
\!\!\!\sqrt{g_{xx}g_{\phi\phi}}\: dx \,d\phi \!=\!
 2\, \frac{\Delta \phi _{\rm N}} {A^2}\,
 \frac{ x_\mathrm{n}-x_\mathrm{s} }
   {(x_\mathrm{n}+\rho)(x_\mathrm{s}+\rho)}\,, \nonumber \\
& &
 \label{area-nariai}
 \end{eqnarray}
where $\sqrt{3\gamma} \leq \rho<\sqrt{6\gamma}$, $x_\mathrm{s}$
and $x_\mathrm{n}$ are defined by (\ref{polos-cold}), and
 $\Delta \phi _{\rm N}$ is
given by (\ref{Period phi}), with $m$ and $q$ subjected to
(\ref{mq-cNariai PCdS}). Thus, ${\cal{A}^{\rm N}}=-8I^{\rm N}$,
where $I^{\rm N}$ is given by (\ref{I-mag-Nariai}), and thus
$\Gamma^{\rm N} \propto e^{S^{\rm N}}$, where $S^{\rm N}=
{\cal{A}^{\rm N}}/4$.
\subsection{\label{sec:Ultracold-S}The ultracold C case. Entropy and area}

In the ultracold  case, the instanton has a single horizon, the
Rindler horizon at $\chi=0$, in its Euclidean section, since
 $\chi=\infty$ is an internal infinity. So, the total area of the
ultracold C instanton is
 \begin{eqnarray}
\!\!\!{\cal{A}}^{\rm u}\!\! =\!\!
 \int_{\chi=0}\!\!\! \sqrt{g_{xx}g_{\phi\phi}}\: dx \,d\phi =\!
 \frac{\Delta \phi _{\rm u}} {A^2}\,
 \frac{ x_\mathrm{n}-x_\mathrm{s} }
   {(x_\mathrm{n}+\rho)(x_\mathrm{s}+\rho)}\,,
 \label{area-ultracold}
 \end{eqnarray}
with $\rho=\sqrt{2(\Lambda+3A^2)}$ [see
(\ref{range-gamma-ultra})], $x_\mathrm{s}$ and $x_\mathrm{n}$ are
defined by (\ref{polos-cold}), and
 $\Delta \phi _{\rm c}$ is
given by (\ref{Period phi}), with $m$ and $q$ subjected to
(\ref{mq-ultra}). It straightforward to verify that ${\cal{A}^{\rm
u}}=-8I^{\rm u}$, where $I^{\rm u}$ is given by
(\ref{I-mag-ultracold}), and thus $\Gamma^{\rm u} \propto
e^{S^{\rm u}}$, where $S^{\rm u}= {\cal{A}^{\rm u}}/4$.

As we have already said, the ultracold C instanton is a limiting
case of both the charged Nariai C instanton and the cold C
instanton (see, e.g., Fig. \ref{mq-fig}). Then, as expected, the
action of the cold C instanton  gives, in this limit, the action
of the ultracold C instanton (see Fig. \ref{cold-fig}). However,
the ultracold frontier of the Nariai C action is given by two
times the ultracold C action (see Fig. \ref{nariai-fig-PCAdS}).
From the results of this section we clearly understand the reason
for this behavior. Indeed, in the ultracold case and in the cold
case, the respective instantons have a single horizon (the other
possible horizon turns out to be an internal infinity). This
horizon gives the only contribution to the total area,
${\cal{A}}$, and therefore to the pair creation rate. In the
Nariai case, the instanton has two horizons with the same area,
and thus the ultracold limit of the Nariai action is doubled with
respect to the true ultracold action.

\subsection{\label{sec:Sub-Maximal-S}The nonextreme sub-maximal case. Entropy and area}

In the lukewarm case, the instanton has two horizons in its
Euclidean section, namely the acceleration horizon at $y=y_A$ and
the black hole horizon at $y=y_+$. So, the total area of the
saddlepoint solution is
\begin{eqnarray}
\!\!\! {\cal A}=\int_{y=y_A}\!\!\! \sqrt{g_{xx}g_{\phi\phi}}\: dx
\,d\phi +\! \int_{y=y_+} \!\!\! \sqrt{g_{xx}g_{\phi\phi}}\: dx
\,d\phi \,,
 \label{area-Sub-Maximal}
 \end{eqnarray}
and once again one has ${\cal{A}}=-8I$, where $I$ is given by
(\ref{I-Sub-Maximal}), and thus $\Gamma \propto e^{S}$, where $S=
{\cal{A}}/4$.

\section{\label{sec:Conc}Summary and discussion}

We have studied in detail the quantum process in which a cosmic
string breaks in a de Sitter (dS) background and a pair of black
holes is created at the ends of the string. The energy to
materialize and accelerate the pair comes from the positive
cosmological constant and from the string tension. This process is
a combination of the processes considered in
\cite{MelMos}-\cite{VolkovWipf}, where the creation of a black
hole pair in a dS background has been analyzed, and in
\cite{HawkRoss-string}-\cite{GregHind}, where the breaking of a
cosmic string accompanied by the creation of a black hole pair in
a flat background has been studied.  We remark that in principle
our explicit values for the pair creation rates also apply to the
process of pair creation in an external electromagnetic field,
with the acceleration being provided in this case by the Lorentz
force instead of being furnished by the string tension. Indeed,
there is no dS Ernst solution, and thus we cannot discuss
analytically the process. However, physically we could in
principle consider an external electromagnetic field that supplies
the same energy and acceleration as our strings and, from the
results of the $\Lambda=0$ case (where the pair creation rates in
the string and electromagnetic cases agree), we expect that the
pair creation rates found in this paper do not depend on whether
the energy is being provided by an external electromagnetic field
or by a string.

We have constructed the saddle point solutions that mediate the
pair creation process through the analytic continuation of the dS
C-metric, and we have explicitly computed the nucleation rate of
the process (see also a heuristic derivation of the rate in the
Appendix). Globally our results state that the dS space is stable
against the nucleation of a string, or against the nucleation of a
string followed by its breaking and consequent creation of a black
hole pair. In particular, we have answered three questions. First,
we have concluded that the nucleation rate of a cosmic string in a
dS background $\Gamma_{\rm string/dS}$ decreases when the mass
density of the string increases. Second, given that the string is
already present in our initial system, the probability
$\Gamma_{\rm BHs/string}$ that it breaks and a pair of black holes
is produced and accelerated apart by $\Lambda$ and by the string
tension increases when the mass density of the string increases.
In other words, a string with a higher mass density makes the
process more probable, for a fixed black hole mass. Third, if we
start with a pure dS background, the probability $\Gamma_{\rm
BHs/dS}$ that a string nucleates on it and then breaks forming a
pair of black holes decreases when the mass density of the string
increases. These processes have a clear analogy with a
thermodynamical system, with the mass density of the string being
the analogue of the temperature $T$. Indeed, from the Boltzmann
factor, $e^{-E_0/(k_{\rm B} T)}$ (where $k_{\rm B}$ is the
Boltzmann constant), one knows that a higher background
temperature turns the nucleation of a particle with energy $E_0$
more probable. However, in order to have a higher temperature we
have first to furnish more energy to the background, and thus the
global process (increasing the temperature to the final value $T$
plus the nucleation of the particle) becomes energetically less
favorable as $T$ increases.

We have also verified that the relation between the rate, entropy
and area, which is satisfied for all the black hole pair creation
processes analyzed so far, also holds in the process studied in
this paper. Indeed, the pair creation rate is proportional to
$e^{S}$, where $S$ is the gravitational entropy of the system, and
is given by one quarter of the total area of all the horizons
present in the saddle point solution that mediates the pair
creation.

To conclude let us recall that the dS C-metric allows two distinct
physical interpretations. In one of them one removes the conical
singularity at the north pole and leaves one at the south pole. In
this way the dS C-metric describes a pair of black holes
accelerated away by a string with positive mass density.
Alternatively, we can avoid the conical singularity at the south
pole and in this case the black holes are pushed away by a strut
(with negative mass density) in between them, along their north
poles. In this paper we have adopted the first choice.
Technically, the second choice only changes the period of the
angular coordinate $\phi$: it would be given by $\Delta
\phi=\frac{4 \pi}{|{\cal G}'(x_\mathrm{s})|}$ instead of
(\ref{Period phi}). We have chosen the first choice essentially
for two reasons. First, the string has a positive mass density
and, in this sense, it is a more physical solution than the strut.
Second, in order to get the above string/pair configuration we
only have to cut the string in a point. The string tension does
the rest of the work. However, if we desire the strut/pair system
described above we would have to cut the strut in two different
points. Then we would have to discard somehow the segment that
joins the black holes along their south poles.


\begin{acknowledgments}

It is a pleasure to acknowledge conversations with Vitor Cardoso
and with Alfredo B. Henriques. This work was partially funded by
Funda\c c\~ao para a Ci\^encia e Tecnologia (FCT) through project
CERN/FIS/43797/2001 and PESO/PRO/2000/4014. OJCD also acknowledges
finantial support from the FCT through PRAXIS XXI programme. JPSL
thanks Observat\'orio Nacional do Rio de Janeiro for hospitality.

\end{acknowledgments}

\appendix*
\section{Heuristic derivation of the nucleation rates}
In order to clarify the physical interpretation of the results, in
this Appendix we heuristically derive the nucleation rates for the
processes discussed in the main body of the paper. We know that an
estimate for the nucleation probability is given by the Boltzmann
factor, $\Gamma \sim e^{-E_0/W_{\rm ext}}$, where $E_0$ is the
energy of the system that nucleates and $W_{\rm ext}=F \ell$ is
the work done by the external force $F$,  that provides the energy
for the nucleation, through the typical distance $\ell$ separating
the created pair.

Forget for a moment the string, and ask what is the probability
that a black hole pair is created in a dS background. This process
has been discussed in \cite{MannRoss} where it was found that the
pair creation rate is $\Gamma \sim e^{-m/\sqrt{\Lambda}}$. In this
case, $E_0 \sim 2m$, where $m$ is the rest energy of the black
hole, and $W_{\rm ext}\sim \sqrt{\Lambda}$ is the work provided by
the cosmological background. To derive  $W_{\rm ext}\sim
\sqrt{\Lambda}$ one can argue as follows. In the dS case, the
Newtonian potential is $\Phi=\Lambda r^2/3$ and its derivative
yields the force per unit mass or acceleration, $\Lambda r$, where
$r$ is the characteristic dS radius, $\Lambda^{-1/2}$. The force
can then be written as $F= {\rm mass}\times{\rm acceleration}\sim
\sqrt{\Lambda}\sqrt{\Lambda}$, where the characteristic mass of
the system is $\sqrt{\Lambda}$. Thus, the characteristic work is
$W_{\rm ext}={\rm force}\times{\rm distance}\sim \Lambda
\Lambda^{-1/2}\sim \sqrt{\Lambda}$, where the characteristic
distance that separates the pair at the creation moment is
$\Lambda^{-1/2}$. So, from the Boltzmann factor we indeed expect
that the creation rate of a black hole pair in a dS background is
given by $\Gamma \sim e^{-m/\sqrt{\Lambda}}$ \cite{MannRoss}.

A question that has been answered in the present paper was: given
that a string is already present in our initial system, what is
the probability that it breaks and a pair of black holes is
produced and accelerated apart by $\Lambda$ and by the string
tension? The presence of the string leads in practice to a problem
in which we have an effective cosmological constant that satisfies
$\Lambda'\equiv \Lambda+3A^2$, that is, the acceleration $A$
provided by the string makes a positive contribution to the
process. Heuristically, we may then apply the same arguments that
have been used in the last paragraph, with the replacement
$\Lambda\rightarrow \Lambda'$. At the end, the Boltzmann factor
tells us that the creation rate for the process is $\Gamma \sim
e^{-m/\sqrt{\Lambda+3A^2}}$. So, for a given black hole mass, $m$,
and for a given cosmological constant, $\Lambda$, the black hole
pair creation process is enhanced when a string is present, as the
explicit calculations done in the main body of the paper show. For
$\Lambda=0$ this heuristic derivation yields $\Gamma \sim
e^{-m/A}$ which is the pair creation rate found in
\cite{HawkRoss-string}.

Another question that we have dealt with in the present paper was:
what is the probability for the nucleation of a string in a dS
background? Heuristically, the energy of the string that nucleates
is $E_0\sim \mu \Lambda^{-1/2}$, i.e., its mass per unit length
times the dS radius, while the work provided by the cosmological
background is still given by $W_{\rm ext}\sim \sqrt{\Lambda}$. The
Boltzmann factor yields for nucleation rate the value $\Gamma \sim
e^{-\mu/\Lambda}$, in agreement with (\ref{PC-rate-stringend}).



\begin{thebibliography}{100}


\bibitem{OscLem_FalVac} O. J. C. Dias,  J. P. S. Lemos,
{\it False vacuum decay: effective one-loop action for pair
creation of domain walls}, J. Math. Phys. {\bf 42}, 3292 (2001);
O. J. C. Dias, in {\it Proceedings of Xth Portuguese Meeting on
Astronomy and Astrophysics}, edited by J. P. S. Lemos et al (World
Scientific, Singapore, 2001), {\tt gr-qc/0106081}.

\bibitem{EQG-book} S. W. Hawking , in {\it Black holes: an Einstein Centenary Survey},
 edited by S. W. Hawking, W. Israel (Cambridge University Press,
 1979); {\it Euclidean Quantum Gravity}, edited by G. W. Gibbons, S. W.
Hawking (Cambridge University Press, 1993).

\bibitem{Ernst}
F. J. Ernst, {\it Removal of the nodal singularity of the
C-metric}, J. Math. Phys. {\bf 17}, 515 (1976).
\bibitem{KW}
W. Kinnersley, M. Walker,  {\it Uniformly accelerating charged
mass in General Relativity}, Phys. Rev. D {\bf 2}, 1359 (1970).
\bibitem{VilenkinStringIpserSikivie} A. Vilenkin, {\it Gravitational
field of vacuum domain walls}, Phys. Lett. B{\bf 133}, 177 (1983);
J. Ipser, P. Sikivie, {\it Gravitationally repulsive domain wall},
Phys. Rev. D{\bf 30}, 712 (1984).

\bibitem{Brown} J. D. Brown, {\it Black
hole pair creation and the entropy factor}, Phys. Rev. D {\bf 51},
5725 (1995);
\bibitem{Melvin} M. A. Melvin,  Phys. Lett. {\bf 8},
65 (1964).

\bibitem{WuSubMax} Z. C. Wu, {\it Quantum creation of a black hole},
 Int. J. Mod. Phys. D {\bf 6}, 199 (1997); {\it Real tunneling and black hole creation},
 Int. J. Mod. Phys. D {\bf 7}, 111 (1998).
\bibitem{BoussoHawkSubMax} R. Bousso,  S. W. Hawking,
{\it Lorentzian condition in quantum gravity}, Phys. Rev. D {\bf
59}, 103501 (1999); {\bf 60}, 109903 (1999) (E).

 \bibitem{Gibbons-book} G. W. Gibbons, in {\it Fields and Geometry},
 Proceedings of the 22nd Karpacz Winter School of Theoretical Physics,
 edited by A. Jadczyk (World Scientific, Singapore, 1986).
\bibitem{GarfGidd}  D. Garfinkle, S. B. Giddings, {\it Semiclassical
Wheeler wormhole production}, Phys. Lett. B{\bf 256}, 146 (1991).
\bibitem{GarfGiddStrom_Sbh} D. Garfinkle, S. B. Giddings, A. Strominger, {\it Entropy in black
hole pair production}, Phys. Rev. D {\bf 49}, 958 (1994)
\bibitem{DGKT} H. F. Dowker, J. P. Gauntlett, D. A. Kastor, J. Traschen,
{\it Pair creation of dilaton black holes}, Phys. Rev. D {\bf 49},
2909 (1994);
\bibitem{DGGH}H. F. Dowker, J. P. Gauntlett, S. B. Giddings, G. T. Horowitz,
{\it Pair creation of extremal black holes and Kaluza-Klein
monopoles}, Phys. Rev. D {\bf 50}, 2662 (1994).
\bibitem{RossU(1)} S. Ross, {\it Pair creation rate for $U(1)^2$ black holes},
Phys. Rev. D {\bf 51}, 2813 (1995).
\bibitem{YiPConeLoop} P. Yi, {\it Toward one-loop tunneling rates of near-extremal
magnetic black hole pair creation}, Phys. Rev. D {\bf 52}, 7089
(1995).
\bibitem{Brown2} J. D. Brown, {\it Duality invariance
of black hole pair creation rates}, Phys. Rev. D {\bf 51}, 5725
(1995).
 \bibitem{HawHorRoss} S. W. Hawking, G. T. Horowitz, S. F. Ross,
 {\it Entropy, area, and black hole pairs},
  Phys. Rev. D {\bf 51}, 4302 (1995).
  \bibitem{HawkHor} S. W. Hawking, G. T. Horowitz,
 {\it The gravitational hamiltonian, action, entropy and surface terms},
  Class. Quant. Grav.  {\bf 13}, 1487 (1996).
\bibitem{Emparan} R. Emparan, {\it Correlations between black holes
 formed in cosmic string breaking},  Phys. Rev. D {\bf 52}, 6976 (1995).

\bibitem{MelMos} F. Mellor, I. Moss, {\it  Black holes and quantum
wormholes}, Phys. Lett. B {\bf 222}, 361 (1989); {\it Black holes
and gravitational instantons}, Class. Quant. Grav. {\bf 6}, 1379
(1989).
  \bibitem{Rom}L. J. Romans, {\it Supersymmetric, cold and lukewarm black
holes in cosmological Einstein-Maxwell theory}, Nucl. Phys. B {\bf
383}, 395 (1992).
\bibitem{MannRoss}
R. B. Mann, S. F. Ross,  {\it Cosmological production of charged
black hole pairs}, Phys. Rev. D {\bf 52}, 2254 (1995).
\bibitem{BoussoHawk} R. Bousso, S. W. Hawking, {\it The probability
for primordial black holes}, Phys. Rev. D {\bf 52}, 5659 (1995);
{\it Pair production of black holes during inflation}, Phys. Rev.
D {\bf 54}, 6312 (1996); S. W. Hawking, {\it Virtual black holes},
Phys. Rev. D {\bf 53}, 3099 (1996).
\bibitem{GaratinniOneLoop} R. Garattini, {\it Energy computation in wormwhole
background with the Wheeler-DeWitt Operators}, Nucl. Phys. B
(Proc. Suppl.) {\bf 57}, 316 (1997); {\it Wormwholes and black
hole pair creation}, Nuovo Cimento B{\bf 113}, 963 (1998);
{Space-time foam, Casimir energy and black hole pair creation},
Mod. Phys. Lett. A {\bf 13}, 159 (1998); {\it Casimir energy and
black hole pair creation in Schwarzschild-de Sitter spacetime},
Class. Quant. Grav. {\bf 18}, 571 (2001).
\bibitem{VolkovWipf} M. Volkov, A. Wipf, {\it Black hole pair creation
in de Sitter space: a complete one-loop analysis}, Nucl. Phys.
B{\bf 582}, 313 (2000).
\bibitem{BooMann} I. S. Booth, R. B. Mann, {\it Complex
instantons and charged rotating black hole pair creation}, Phys.
Rev. Lett. {\bf 81}, 5052 (1998); {\it Cosmological pair
production of charged and rotating black holes}, Nucl. Phys. B
{\bf 539}, 267 (1999).
 \bibitem{BoussoDil} R. Bousso, {\it Charged Nariai black holes with a
dilaton}, Phys. Rev. D {\bf 55}, 3614 (1997).

\bibitem{HawkRoss-string} S. W. Hawking, S. F. Ross, {\it Pair production of
black holes on cosmic strings}, Phys. Rev. Lett. {\bf 75}, 3382
(1995).
 \bibitem{DougHorKastTras}  D. M. Eardley G. T.
Horowitz, D. A. Kastor, J. Traschen, {\it Breaking cosmic strings
without black holes}, Phys. Rev. Lett. {\bf 75}, 3390 (1995).
\bibitem{AchGregKui} A. Ach\'ucarro, R. Gregory, K. Kuijken, {\it Abelian Higgs
 hair for black holes}, Phys. Rev. D {\bf 52}, 5729 (1995).
 \bibitem{GregHind}   R. Gregory, M. Hindmarsh, {\it Smooth metrics
for snapping strings}, Phys. Rev. D {\bf 52}, 5598 (1995).
 \bibitem{PreskVil} J. Preskill, A. Vilenkin, {\it Decay of metastable topological defects},
 Phys. Rev. D {\bf 47}, 2324 (1993).
\bibitem{Empar-string} R. Emparan, {\it Pair production
of black holes joined by cosmic strings}, Phys. Rev. Lett. {\bf
75}, 3386 (1995).

\bibitem{CaldChamGibb} R. R. Caldwell, A. Chamblin, G. W. Gibbons,
{\it Pair creation of black holes by domain walls}, Phys. Rev. D
{\bf 53}, 7103 (1996).
\bibitem{BouCham}  R. Bousso, A. Chamblin,
{\it Patching up the no-boundary proposal with virtual Euclidean
wormholes}, Phys. Rev. D {\bf 59}, 084004 (1999).



\bibitem{MannAdS} R. Mann, {\it Pair production of topological
anti-de Sitter black holes}, Class. Quantum Grav. {\bf 14}, L109
(1997); {\it Charged topological black hole pair creation}, Nucl.
Phys. B {\bf 516}, 357 (1998).

\bibitem{Other} R. Parentani, S. Massar, {\it The Schwinger mechanism, the Unruh effect
and the production of accelerated black holes}, Phys. Rev. D {\bf
55}, 3603 (1997); Z. C. Wu, {\it Pair creation of  black holes in
anti-de Sitter space background (I)}, Gen. Rel. Grav. {\bf 31},
223 (1999); {\it Pair creation of black hole in anti-de Sitter
space background},
 Phys. Lett. B {\bf 445}, 274 (1999); {\it Quantum creation of topological black hole},
 Mod. Phys. Lett. A {\bf 15}, 1589 (2000); J. Garriga, M. Sasaki,
 {\it Brane-world creation and black holes}, Phys. Rev. D {\bf
62}, 043523 (2000); R. Emparan, R Gregory, C. Santos,
 {\it Black holes on thick branes}, Phys. Rev. D {\bf
63}, 104022 (2001).


\bibitem{HawkRoss}
S. W. Hawking, S. F. Ross,  {\it Duality between electric and
magnetic black holes}, Phys. Rev. D {\bf 52}, 5865 (1995).

\bibitem{InformationLossLOSS}  S. W. Hawking, {\it Breakdown of predictability
in gravitational collapse}, Phys. Rev. D{\bf 14}, 2460 (1976).
\bibitem{InformationLossREVIEW} T. Banks, {\it Lectures on black holes and information loss},
 {\tt hep-th/9412131};  S. B. Giddings, {\it The black hole information
paradox}, {\tt hep-th/9508151}; {\it Why aren't black holes
infinitely produced?},  Phys. Rev. D {\bf 51}, 6860 (1995); L.
Susskind, {\it Trouble for remnants}, {\tt hep-th/9501106}.

\bibitem{PlebDem}
 J. F. Pleba\'nski, M. Demia\'nski, {\it Rotating, charged and
uniformly accelerating mass in general relativity},  Annals of
Phys. (N.Y.) {\bf 98},  98 (1976).
\bibitem{PodGrif2} J. Podolsk\'y,  J.B. Griffiths,
  {\it Uniformly accelerating black holes in a de Sitter universe},
  Phys. Rev. D  {\bf 63}, 024006 (2001).
\bibitem{OscLem_dS-C}  O. J. C. Dias, J. P. S. Lemos,
{\it Pair of accelerated black holes in a de Sitter background:
the dS C-metric}, Phys. Rev. D {\bf 67}, 084018 (2003).

\bibitem{VitOscLem} V. Cardoso, O. J. C. Dias, J.
P. S. Lemos, {\it Gravitational radiation in D-dimensional
spacetimes}, Phys. Rev. D {\bf 67}, 064026 (2003).
\bibitem{BPP}
J. Bi\v c\'ak, {\it Gravitational radiation from uniformly
accelerated particles in general relativity},
 Proc. Roy. Soc.  A {\bf 302}, 201 (1968);
 V. Pravda, A. Pravdova, {\it Boost-rotation symmetric spacetimes
- review}, Czech. J. Phys. {\bf 50}, 333 (2000); {\it  On the
spinning C-metric}, in {\it Gravitation: Following the Prague
Inspiration}, edited by O. Semer\'ak, J. Podolsk\'y, M. Zofka
(World Scientific, Singapore, 2002), {\tt gr-qc/0201025}.
\bibitem{BicKrt}
J. Bi\v c\'ak, P. Krtou\v s, {\it Accelerated sources in de Sitter
spacetime and the insuffiency of retarded fields}, Phys. Rev. D
{\bf 64}, 124020 (2001); {\it The fields of uniformly accelerated
charges in de Sitter spacetime}, Phys. Rev. Lett. {\bf 88}, 211101
(2002).
\bibitem{KrtPod}
P. Krtou\v s, J. Podolsk\' y, {\it Radiation from accelerated
black holes in de Sitter universe}, {\tt gr-qc/0301110}.


\bibitem{HartleHawk} J.B. Hartle, S.W. Hawking, {\it Wave function of the
Universe}, Phys. Rev. D {\bf 28}, 2960 (1983).

 \bibitem{GibbHawPer}  G. W. Gibbons, S. W. Hawking,  M. J.
Perry, {\it Path integral and the indifiniteness of the
gravitational action},
 Nucl. Phys. B {\bf 138}, 141 (1978).
 \bibitem{GrossPerryYaffe} D. J. Gross, M. J. Perry, L. G. Yaffe, {\it Instability
of flat space at finite temperature}, Phys. Rev. D {\bf 25}, 330
(1982).
\bibitem{GibbPer}  G. W. Gibbons, M. J. Perry,
 {\it Quantizing gravitational instantons},
 Nucl. Phys. B {\bf 146}, 90 (1978).
\bibitem{ChristDuff} S. M. Christensen, M. J. Duff,
 {\it Quantizing gravity with cosmological constant},
 Nucl. Phys. B {\bf 146}, 90 (1978).
\bibitem{GinsPerry} P. Ginsparg, M. J. Perry, {\it Semiclassical
perdurance of de Sitter space}, Nucl. Phys. B {\bf 222}, 245
(1983).
\bibitem{Young} R. E. Young, {\it Semiclassical stability of
asymptotically locally flat spaces}, Phys. Rev. D {\bf 28}, 2420
(1983).
\bibitem{YoungdS} R. E. Young,  {\it Semiclassical instability of gravity with positive
cosmological constant}, Phys. Rev. D {\bf 28}, 2436 (1983).

\bibitem{BrownYork} J. D. Brown, J. W. York, {\it Quasilocal energy and
conserved charges derived from the gravitational action}, Phys.
Rev. D {\bf 47}, 1407 (1993); , {\it Microcanonical functional
integral for the gravitational field}, Phys. Rev. D {\bf 47}, 1420
(1993).

\bibitem{GibbHawk} G. W. Gibbons, S. W. Hawking, {\it Action integrals and
partition functions in quantum gravity}, Phys. Rev. D {\bf 15},
2752 (1977).

\bibitem{OscLem_AdS-C}  O. J. C. Dias, J. P. S. Lemos,
{\it Pair of accelerated black holes in an anti-de Sitter
background: the AdS C-metric}, Phys. Rev. D {\bf 67}, 064001
(2003).

\bibitem{OscLem_nariai}  O. J. C. Dias, J. P. S. Lemos,
{\it The extremal limits of the C-metric: Nariai,
Bertotti-Robinson and anti-Nariai C-metrics}, accepted for
publication in  Phys. Rev. D (2003), {\tt hep-th/0306194}.

\bibitem{Bousso60y} S. Nojiri, S. D. Odintsov, {\it Effective action for
conformal scalars and anti-evaporation of black holes}, Int. J.
Mod. Phys. A {\bf 14}, 1293 (1999); {\it Quantum evolution of
Schwarzschild-de Sitter (Nariai) black holes}, Phys. Rev. D {\bf
59}, 044026 (1999); R. Bousso, {\it Proliferation of de Sitter
space}, Phys. Rev. D {\bf 58}, 083511 (1998); {\it Adventures in
de Sitter space}, {\tt hep-th/0205177}.



\bibitem{ReggeGibbonsPerryAconSing} T. Regge, Nuovo Cimento {\bf 19},
558 (1961); G. W. Gibbons, M. J. Perry, Phys. Rev. D {\bf 22}, 313
(1980).


\end{thebibliography}
\end{document}